\pdfoutput=1
\newcommand*{\ATLASLATEXPATH}{latex/}
\documentclass[UKenglish,texlive=2011,cernpreprint,PAPER,txfonts]{\ATLASLATEXPATH atlasdoc}
\usepackage[subfigure=true]{\ATLASLATEXPATH atlaspackage}
\usepackage{caption}
\usepackage{\ATLASLATEXPATH atlasbiblatex}

\usepackage{\ATLASLATEXPATH atlascontribute}

\usepackage{\ATLASLATEXPATH atlasphysics}
\usepackage{amsmath}
\usepackage{bm}
\usepackage{siunitx}
\usepackage{multirow}
\usepackage{tabularx}
 
\graphicspath{{logos/}{figures/}}

\usepackage{dbcomb-defs}

\AtlasTitle{Searches for heavy diboson resonances in $pp$ collisions
at $\sqrt{s}=13$~\TeV{} with the ATLAS detector}

\AtlasRefCode{EXOT-2016-01}

\PreprintIdNumber{CERN-EP-2016-106}

\AtlasJournalRef{JHEP 09 (2016) 173}
\AtlasDOI{doi:10.1007/JHEP09(2016)173}

\AtlasAbstract{Searches for new heavy resonances decaying to $WW$, $WZ$,
and $ZZ$ bosons are presented, using a data sample corresponding to 3.2 fb$^{-1}$
of $pp$ collisions at $\sqrt{s}=13$~\TeV{} collected with the ATLAS
detector at the CERN Large Hadron Collider.  Analyses selecting
\vvqq, \lvqq, \llqq{} and \qqqq{} final states are
combined, searching for a narrow-width resonance with mass between 500
and 3000~\GeV. The discriminating variable is either an invariant mass
or a transverse mass. No significant
deviations from the Standard Model predictions
are observed.
Three benchmark models are tested:  a model predicting the existence of a new heavy scalar singlet, a simplified model predicting a
heavy vector-boson triplet, and a bulk Randall-Sundrum model with a heavy spin-2
graviton.
Cross-section limits  are set at the 95\% confidence level and are
compared to theoretical cross-section predictions for a variety of models. 
The data exclude a scalar singlet with mass below \limScal~\GeV, a heavy vector-boson 
triplet with mass below \limHvt~\GeV{},  and a 
graviton with mass below \limGrav~\GeV.
These results significantly extend the previous limits set using $pp$ collisions at $\sqrt{s}=8\tev$.
}

\newcommand{\muhat}{\ensuremath{\hat{\mu}}}

\begin{document}

\maketitle

\tableofcontents

\section{Introduction}
\label{sec:introduction}

Diboson resonances are predicted in several extensions to the Standard Model (SM), such
as composite Higgs models~\cite{Dugan:1984hq,Agashe:2004rs}, 
technicolour~\cite{Eichten:2007sx,Sannino:2004qp,Andersen:2011yj},
warped extra dimensions~\cite{Randall:1999ee,Randall:1999vf,Davoudiasl:2000wi},
Two-Higgs-doublet models (2HDM)~\cite{Branco:2011iw}, and Grand Unified
Theories~\cite{Pati:1974yy,Georgi:1974sy,Georgi:1974my,Fritzsch:1974nn}. 
The search for high-mass resonances decaying into vector bosons
benefits greatly from the
increase in centre-of-mass energy of
proton--proton collisions at the Large Hadron Collider (LHC)
from $\sqrt{s}=8\tev$ (Run~1) to $13\tev$ (Run~2).
This would result in more abundant production of new particles with masses 
significantly in excess of a \tev,
in processes initiated by $gg$, $gq$ or
$qq$.\footnote{To simplify notation, antiparticles are denoted by the same symbol as the corresponding particles.}  

This paper reports a search for a charged or neutral
resonant state, with a mass
between 500 \gev{} and 3 \tev, decaying to $WW$, $ZZ$ or $WZ$ bosons, with subsequent decays of the $W$ and $Z$ bosons to quarks
or leptons. Four different decay modes are considered:
the fully hadronic mode (\qqqq), and the semileptonic modes (\llqq, \lvqq{}  and \vvqq). Decays of the $W$ or $Z$ bosons to quarks are 
reconstructed as single jets with a large radius parameter.
These jets are required to have  features characteristic of a two-body decay, and are identified as $W$ or
$Z$ bosons using the jet mass and jet substructure~\cite{ATL-PHYS-PUB-2015-033,CERN-PH-EP-2015-204}. 

Three specific signal models are used to assess the sensitivity of the search, to optimise the event selection,
and to search for local excesses in the observed data.  
The first is an extension of the SM with an additional
 heavy, CP-even, scalar singlet decaying to longitudinally polarised
 bosons~\cite{dilaton}.
The second is the  Heavy Vector Triplet (HVT) parameterisation~\cite{Pappadopulo:2014qza}, predicting $W'\rightarrow WZ$ and $Z' \rightarrow WW$ processes.
The third model, known as a bulk Randall--Sundrum (RS) graviton model, features a spin-2
graviton ($\grav$) decaying to $WW$ or $ZZ$. The $\grav$ is the first 
Kaluza--Klein mode in a RS
model~\cite{Randall:1999ee,KKmodesHLZ} with a warped extra dimension 
with curvature $\kappa$, where the SM fields are allowed to propagate
in the bulk of the extra dimension ~\cite{ADPS2007,AAS2008,AS2008}. 

Both ATLAS and CMS have searched for heavy diboson resonances in various
final states in the $\sqrt{s} = 7\tev$ and
8~\TeV\ datasets~\cite{Aad:2014pha,Aad:2014xka,Aad:2015ufa,Aad:2015owa,Khachatryan:2014hpa,Khachatryan:2014gha,Khachatryan:2014xja,Aad:2015kna,Aad:2015agg,Aad:2015ipg}.
As an example, CMS set a lower limit of $1.7\tev$ at the 95\% confidence level (CL) on the mass of a $W'$ boson predicted by an Extended Gauge Model
(EGM)~\cite{EGM} using the fully hadronic channel~\cite{Khachatryan:2014hpa}.
The \qqqq, \llqq, \lvqq{}  channels were combined by
ATLAS using the bulk RS \grav{} model and the EGM $W'$  boson as benchmarks~\cite{Aad:2015ipg}. Observed lower limits at 95\% CL of $1.81\tev$ on the EGM $W'$ mass and of $810\gev$ on the bulk \grav{} mass were obtained, assuming 
$\kappa/\bar{M}_\text{Pl}=1$
(where
 $\bar{M}_\text{Pl}$ is the reduced Planck mass) for the bulk \grav{} signal hypothesis.
The largest deviation from the predicted background in that analysis
was a 2.5$\sigma$ local excess  close to a mass of 2~\TeV.

\section{ATLAS detector and data sample}
\label{sec:atlas-data}

The ATLAS detector~\cite{Aad:2008zzm} is a general-purpose particle
detector used to investigate a broad range of physics processes.
It includes inner tracking devices surrounded by a
superconducting solenoid, electromagnetic (EM) and hadronic calorimeters,
and a muon spectrometer  inside a system of toroid magnets.
The inner detector (ID) consists of a silicon pixel detector including
the newly installed 
Insertable B-Layer~\cite{ibl}, a silicon microstrip
detector and a straw tube tracker. It is situated inside a 2 T axial magnetic
field from the solenoid and provides precision tracking of charged
particles with pseudorapidity\footnote{ATLAS uses a right-handed coordinate system with its origin
  at the nominal interaction point (IP) in the centre of the detector
  and the $z$-axis along the beam pipe. The $x$-axis points from the
  IP to the centre of the LHC ring, and the $y$-axis points
  upwards. Cylindrical coordinates $(r,\phi)$ are used in the
  transverse plane, $\phi$ being the azimuthal angle around the beam
  pipe.  The pseudorapidity is defined in terms of the polar angle
  $\theta$ as $\eta = -\ln \tan(\theta/2)$.  Rapidity is also defined
  relative to the beam axis as $y=0.5\ln [(E+p_z)/(E-p_z)]$.
  Angular distance is
  measured in units of $\Delta R \equiv \sqrt{(\Delta\eta)^{2} +
    (\Delta\phi)^{2}}$.}  $|\eta| < 2.5$. The straw tube tracker
also provides transition radiation measurements for electron
identification. The calorimeter system covers the pseudorapidity range
$|\eta| < 4.9$. It is composed of sampling calorimeters with either
liquid argon or scintillator tiles as the active medium. The muon
spectrometer (MS) provides muon identification and  measurement for
$|\eta| < 2.7$  and 
detectors for triggering in the region $|\eta|<2.4$. The ATLAS detector has a two-level trigger system to
select events for offline analysis~\cite{ATL-DAQ-PUB-2016-001}. 

The data used in this analysis were recorded with the ATLAS detector during the 2015 run and correspond to an integrated luminosity of $3.2\pm 0.2$ fb$^{-1}$ of proton--proton collisions at $\sqrt{s}=$ 13 TeV. The measurement of the integrated luminosity is derived, following
a methodology similar to that detailed in Ref.~\cite{ATLAS_luminosity}, 
from a preliminary calibration of the luminosity scale using $x$–$y$ beam-separation scans performed in August 2015. The data are required to satisfy a number of conditions ensuring that the detector was operating well
while the data were recorded.

\section{Signal and background simulation}
\label{sec:simulation}

The Monte Carlo (MC) simulation of three benchmark signal models is used to optimise the sensitivity of the search and to interpret the results.

The first model extends the SM by adding a new, heavy, neutral Higgs boson,
using  the narrow-width approximation
(NWA) benchmark~\cite{PhysRevD.36.3463,Barger:2007im}.
Results are then interpreted within a model of a CP-even scalar singlet $S$~\cite{dilaton}.  The model is parameterised by: an energy scale $\Lambda=1~\TeV$; a coefficient $c_H$ scaling the coupling of $S$ to the Higgs boson; and a coefficient $c_3$ scaling the coupling of $S$ to gluons.  Two benchmark scenarios are considered, one in which $c_3$ is set via \emph{naive dimensional analysis} (NDA) to be $c_3=(1/4\pi)^2$, with $c_H=0.9$; and another in which the coupling to gluons is \emph{Unsuppressed} and $c_3=1/8\pi$, with $c_H=0.5$.  The value of $c_3$ determines the production cross-section and the decay width to gluons, while 
decays to $W$ or $Z$ bosons account for the remaining decay width.
In the Unsuppressed scenario considered in this paper, the total branching ratio to $WW$, $ZZ$ or $HH$ increases from 59\% at 500~\gev, to 70\% at 2~\TeV\ and to 73\% at 5~\TeV. 
For the NDA scenario, this branching ratio is always above 99\%.  The ratio of $WW$:$ZZ$:$HH$ partial widths is approximately $2$:$1$:$1$ in both scenarios, and couplings to fermions and transversely polarised bosons are set to zero.

The second model is based on the HVT phenomenological Lagrangian which
introduces a new triplet of heavy vector bosons that contains three
states with identical masses: the two electrically charged $W'$ bosons and
the electrically neutral $Z'$ boson. The Lagrangian parameterises the
couplings of the new HVT with the SM fields in a generic manner. This
parameterisation allows a large class of models to be described, in which
the new triplet field mixes with the SM vector bosons. The coupling
between the new triplet and the SM fermions is given by the
combination of parameters $g^2 c_F / g_V$, where $g$ is the SM
$SU(2)_{\mathrm L}$ gauge coupling, $c_F$ is a multiplicative factor that
modifies the coupling to fermions, and $g_V$ represents the coupling
strength of the known $W$ and $Z$ bosons to the new vector bosons. Similarly, the coupling between the
Higgs boson and the new triplet is given by the combination $g_V c_H$,
where $c_H$ is a multiplicative factor that modifies the coupling to
the Higgs boson.
Other coupling parameters involving more than one heavy vector boson
give negligible contributions to
the overall cross-sections for the processes of interest here.
Two benchmarks are used~\cite{Pappadopulo:2014qza}. 
In the first one, referred to as \textit{model-A} with $g_V =
1$, the branching ratios of the new HVT
to fermions and gauge bosons are similar to those predicted by some
extensions of the SM gauge group~\cite{Barger:1980ix}.  This model, although severely constrained by searches for new resonances decaying to 
leptons~\cite{Aad:2014cka,ATLAS:2014wra,Khachatryan:2014fba,Khachatryan:2014tva}, is included because of its similarity to the EGM $W'$ model used as a benchmark in previous searches~\cite{Aad:2015ipg}.
In the second
model, referred to as \textit{model-B} with $g_V = 3$, the fermionic
couplings of the new HVT 
are suppressed, and branching ratios are similar to the ones
predicted by composite  Higgs boson models~\cite{Contino:2011np,Bellazzini:2014yua,Panico:2015jxa}. In both
benchmarks the width of the HVT is narrower than the detector
resolution, and the kinematic distributions relevant to this search are
very similar. Off-shell and interference effects are not considered. 

The third model considered is the so-called bulk RS
model~\cite{ADPS2007}. This model extends the original RS
model with one warped extra
dimension~\cite{Randall:1999ee,Randall:1999vf} by  
allowing the SM fields to propagate in the bulk of the extra dimension.  This avoids constraints on the original RS model from limits
	on flavour-changing neutral currents and electroweak precision measurements~\cite{graviton}.
	This model is characterised by the dimensionless coupling constant
	$\kappa/\bar{M}_{{\text{Pl}}}\sim\mathcal{O}(1)$.
	In this model the branching ratio of the Kaluza--Klein graviton (\grav{})  to pairs of vector bosons, $WW$ or $ZZ$,  is larger than 30\%.

For the NWA Higgs boson model, samples are generated for gluon fusion
production with QCD corrections up to next-to-leading order (NLO), assuming a Higgs boson decay width of $4\MeV$.
Events are generated using \textsc{POWHEG BOX}~\cite{powheg3} v1 r2856 
with the CT10 parton distribution function (PDF) set~\cite{Lai:2010vv} interfaced to \textsc{Pythia} 8.186~\cite{Sjostrand:2007gs} using the AZNLO~\cite{Aad:2014xaa} tune of the underlying event.

Benchmark  samples of the HVT and bulk RS  graviton are
generated using \textsc{MadGraph5\_aMC@NLO} 2.2.2~\cite{Alwall:2014hca}
interfaced to \textsc{Pythia} 8.186 with the NNPDF23LO PDF set~\cite{Ball:2012cx}
for resonance masses ranging from 0.5~\tev{} to 5~\tev.  
For the HVT interpretation, samples are generated according to model A, 
for resonance masses ranging from 0.5~\tev{} to 3~\tev{} for the semileptonic channels and from 1.2~\tev{} to 3~\tev{} for the fully hadronic search.
Interpretation in the HVT model-B, $g_V=3$ scenario uses the model A signal samples rescaled to the predicted cross-sections from model-B.
For the bulk RS graviton model, the curvature scale parameter $\kappa/\bar{M}_\text{Pl} $ is assumed to be 1. 
Table \ref{tab:simgravhvt} shows the resonance width and the product of cross-sections and branching ratios for the various models.

\begin{table}[!pht]
\caption{The resonance width ($\Gamma$) and the product of
  cross-section times branching ratio ($\sigma\times\text{BR}$) for diboson final states, for different values of the pole mass $m$ of the resonances for a representative benchmark for the spin-$0$, spin-$1$ and spin-$2$ cases. The table shows the predictions by the CP-even scalar model ($\Lambda=1~\TeV$, $c_H=0.9$, $c_3=1/16\pi^2$), by model-B of the HVT parameterisation ($g_V = 3$), and by the graviton model ($\kappa/\bar{M}_\text{Pl} = 1$).
  In the case of the scalar and HVT models, the alternate benchmarks (Unsuppressed scenario, model-A) correspond to a different cross-section but similar resonance width and ratios between the branching ratios into $WW/WZ/ZZ$.
}\label{tab:simgravhvt}
\begin{center}
\begin{tabular}{c|SSS|SSS|SSS}
  \hline\hline
  & \multicolumn{3}{c|}{Scalar}
  & \multicolumn{3}{c|}{HVT $W'$ and $Z'$}
  & \multicolumn{3}{c}{\grav} \\
  \cline{2-10}
  & & $WW$ & $ZZ$ & & $WW$ & $WZ$ & & $WW$ & $ZZ$ \\
  $m$
  &    $\Gamma$  & $\sigma\times\text{BR}$ & $\sigma\times\text{BR}$
  &    $\Gamma$  & $\sigma\times\text{BR}$ & $\sigma\times\text{BR}$ 
  &    $\Gamma$  & $\sigma\times\text{BR}$ & $\sigma\times\text{BR}$  \\
  $[\text{TeV}]$
  & $[\text{\GeV}]$  & [fb]   &[fb]
  & $[\text{\GeV}]$  & [fb]   &[fb]
  & $[\text{\GeV}]$  & [fb]   &[fb] \\
\hline
  0.8 &  3.9 & 37   & 18   & 32  & 354  & 682   & 46  & 301  & 155 \\ 
 1.6 &  33  & 2.5  & 1.3  & 51  & 38.5 & 79.3  & 96  & 4.4  & 2.2 \\
 2.4 &  111 & 0.32 & 0.16 & 74  & 4.87 & 10.6  & 148 & 0.28 & 0.14\\
\hline\hline
\end{tabular} 
\end{center}
\end{table}

MC samples are used to model the shape and normalisation of the relevant kinematic distributions for most SM background processes in the \vvqq{}, \lvqq{} and \llqq{} searches.
Events containing $W$ or $Z$ bosons with associated jets are simulated
using the \textsc{Sherpa} 2.1.1~\cite{Gleisberg:2008ta} generator. Matrix
elements (ME) are calculated for up to two partons at NLO and four
partons at leading order (LO) using the Comix~\cite{Gleisberg:2008fv}
and OpenLoops~\cite{Cascioli:2011va} ME generators. They are merged with the
\textsc{Sherpa} parton shower (PS)~\cite{Schumann:2007mg} using the
ME+PS@NLO prescription~\cite{Hoeche:2012yf}. The CT10 PDF
set is used in conjunction with a dedicated set of
tuned parton-shower parameters developed by the \textsc{Sherpa} authors.
For the generation of top--antitop pairs (\ttbar) and single top-quarks in the $Wt$- and
$s$-channels the \textsc{POWHEG BOX} v2~\cite{powheg1,powheg2,powheg3}
generator with the CT10 PDF set
is used. Electroweak ($t$-channel) single-top-quark events are generated
using the \textsc{POWHEG BOX} v1 generator. This generator uses the
four-flavour scheme for the NLO ME calculations together
with the four-flavour PDF set CT10f4~\cite{Lai:2010vv}. For all top-quark processes,
top-quark spin correlations are preserved; for $t$-channel production, top-quarks
are decayed using \textsc{MadSpin}~\cite{Artoisenet:2012st}.  The
parton shower, fragmentation, and the underlying event are simulated
using \textsc{Pythia} 6.428~\cite{Sjostrand:2006za} with the
CTEQ6L1\cite{1126-6708-2002-07-012} PDF sets and the set of tuned
parameters known as the ``Perugia 2012
tune''~\cite{Skands:2010ak}. The top-quark mass is assumed to be 
172.5 GeV. The \textsc{EvtGen} v1.2.0 program~\cite{EvtGen} is used for
the bottom- and charm-hadron decays.

The cross-sections calculated at next-to-next-to-leading order (NNLO) accuracy  for
$W/Z$+jets~\cite{Melnikov:2006kv} and at NNLO+NNLL (next-to-next-to-leading-logarithm) accuracy for
\ttbar\ production~\cite{Czakon:2013goa} are used to normalise the samples for the optimisation studies, but the final normalisations of these dominant backgrounds are determined by fitting kinematic distributions to the data. For single-top-quark production, cross-sections are taken from Ref.~\cite{Kidonakis:2011wy}.

Diboson processes with one boson decaying hadronically and the
other decaying leptonically are simulated using the \textsc{Sherpa} 2.1.1
generator. They are calculated for up to one ($ZZ$) or no ($WW$, $WZ$)
additional partons at NLO, and up to three additional partons at LO using
the Comix and OpenLoops ME generators. They are merged with
the \textsc{Sherpa} PS using the ME+PS@NLO
prescription. The CT10 PDF set is used in conjunction with a dedicated
parton-shower tuning developed by the \textsc{Sherpa} authors. 
Cross-section values from the generator, which are at NLO accuracy, are used.

The dominant background in the fully hadronic  final state is from multi-jet events.
While the background in this search is estimated directly from data, samples of simulated dijet events 
are produced, using
\textsc{Pythia} 8.186 with the
NNPDF23LO PDFs and the parton-shower parameter set known as the ``A14 tune''~\cite{ATL-PHYS-PUB-2014-021},
to characterise the invariant mass distribution of the dijet final state and optimise the sensitivity of the search.
The \textsc{EvtGen} v1.2.0 program is used for
the bottom- and charm-hadron decays.

All simulated MC samples include the effect of multiple proton--proton interactions in the same and neighbouring bunch crossings (pile-up) by overlaying simulated minimum-bias events, generated with \textsc{Pythia} 8.186, on each generated signal or background event.  The generated samples are processed through the \textsc{Geant4}-based ATLAS detector simulation~\cite{AGO-0301,ATLASSIM}.
Simulated events are reconstructed with the standard ATLAS reconstruction software used for collision data. Table~\ref{tab:sim_background} summarises the background MC samples used.

\begin{table}[!htbp]
\centering
\caption{Generators and PDFs used in the simulation of the various background processes.}
\label{tab:sim_background}
\begin{tabular}{l|l|l}
\hline
Process                                & PDF          & Generator          \\  \hline                       
\multirow{1}{*}{$W/Z$ + jets} & CT10              & \textsc{Sherpa} 2.1.1   \\                              
\multirow{1}{*}{$tt$}        & CT10                      &
\textsc{POWHEG BOX} v2+\textsc{Pythia} 6.428 \\                               
\multirow{1}{*}{Single top-quark ($Wt$, $s$-channel)}        &
CT10                    &  \textsc{POWHEG BOX} v2+\textsc{Pythia} 6.428 \\       
\multirow{2}{*}{Single top-quark ($t$-channel)} & \multirow{2}{*}{CT104f        }
&  \textsc{POWHEG BOX} v1+\textsc{Pythia} 6.428\\
&  &  + \textsc{MadSpin} 2.1.2 \\       
\multirow{1}{*}{Diboson ($WW$, $WZ$, $ZZ$)}                  & CT10
& \textsc{Sherpa} 2.1.1     \\  
\multirow{1}{*}{Dijet}                  & NNPDF23LO                    & \textsc{Pythia} 8.186     \\   \hline                                           
\end{tabular}
\end{table}

\section{Object reconstruction and selection}
\label{sec:objects}

Electrons are reconstructed from clusters of energy deposits in the
EM calorimeter that match a track reconstructed in the ID.
The electrons used are required to
have transverse momentum $\pT > 7$~\GeV\ and $|\eta|<2.47$. 
They are
identified using a likelihood identification criterion described in
Ref.~\cite{ATLAS-CONF-2014-032}.
 The levels of identification are
categorised as ``loose'', ``medium'' and ``tight'', which
correspond to approximately $96\%$, $94\%$ and $88\%$
identification efficiency for an electron with transverse energy
(\et) of $100\,\gev$, where \et is defined in terms of the energy $E$ and of the polar angle $\theta$ as $\et=E\sin\theta$.

Muons are reconstructed by combining ID and MS tracks. They are classified as
``medium'' if they satisfy identification requirements based on the
number of hits in the different ID and MS subsystems and on the
compatibility of track curvature measurements in the two
subsystems~\cite{Aad:2016jkr}.
An additional sample of ``loose'' muons is constructed including all
medium muons,  muons identified by combining an ID track with
at least one track segment reconstructed in the MS, and muons 
reconstructed in the $|\eta|<0.1$ region, where
the MS is lacking coverage, by associating
an ID track to an energy
deposit in the calorimeters compatible with a minimum-ionising
particle.  Muons are required to have $\pt>7\gev$ and $|\eta|<2.7$.   
The loose and medium muons have average efficiencies of about 98\% and 96\% for $|\eta|<2.5$, respectively.

In order to ensure that leptons originate from the
interaction point, requirements of $|d_0^\text{BL}|/\sigma_{d_0^\text{BL}}<5\,(3)$ and
$|z_0^\text{BL} \sin\theta|<0.5$~mm are imposed on the tracks associated with the electrons (muons), where
$d_0^\text{BL}$ is the transverse impact parameter of the
track with respect to the measured beam line (BL) position
determined at the point of closest approach of the track to the beam line, 
$\sigma_{d_0^\text{BL}}$ is the uncertainty in the measured
$d_0^\text{BL}$, $z_0^\text{BL}$ is the difference between the
longitudinal position of the track along the beam line at the point
where $d_0^\text{BL}$ is measured and the longitudinal position of the
primary interaction vertex,\footnote{If more than one vertex is reconstructed, the one
with the highest sum of $\pt^2$ of the associated tracks is regarded
as the primary vertex.} and $\theta$ is the polar angle of the track.
Lepton isolation criteria are defined based on low values for the scalar sum of
transverse momenta of tracks with $\pt>1\GeV$ within a $\Delta R$ cone
around the lepton, whose size depends upon its \pt, and excluding the track
associated with the lepton (\emph{track isolation}). These criteria are
optimised for a uniform efficiency of $99\%$ in the $(\pt,\eta)$ plane
for leptons from $Z\to\ell\ell$ decays in \Zjet{}
events.  Calorimeter isolation is also used for the $\lvqq$ channel, using an isolation variable
constructed from calorimeter activity within a cone of radius $\Delta R = 0.2$ around the lepton candidate.
The isolation criteria depend on both \pt{} and \eta, and accept 95\% of $Z\to\ell\ell$ events while
maximising the rejection of leptons originating in jets.

Jets are reconstructed from three-dimensional topological
clusters of energy deposits in the calorimeter calibrated at the EM scale~\cite{Aad:2016upy}, using the anti-$k_t$ 
algorithm~\cite{antikt} with two different radius 
 parameters of $R=1.0$ and $R=0.4$, hereafter referred to as large-$R$ jets (denoted
by ``$J$'') and small-$R$ jets (denoted by ``$j$''), respectively.
The four-momenta of the jets are calculated as the sum of the
four-momenta of the clusters, which are assumed to be massless.

The \pT\  of small-$R$ jets are
corrected for losses in passive material, the non-compensating
response of the calorimeter, and contributions from
pile-up~\cite{Aad:2011he}. They are required to have $\pT>20\,\GeV$
and $|\eta|<2.4$.
For small-$R$ jets with $\pT<50\,\GeV$, 
a jet vertex tagger (JVT)~\cite{JVT} discriminant, based on tracking and vertexing information, is
required to be larger than $0.64$, where
the JVT is a multivariate tagger used to identify and remove  jets
with a large contribution from pile-up.
In addition, small-$R$ jets are discarded if they are within a cone of size
 $\Delta R<0.2$  around an electron
candidate, or if they have less than three associated tracks and are within a cone of size
 $\Delta R<0.2$ around a muon candidate. However, if a small-$R$ jet with three or more associated 
 tracks is within a cone of size $\Delta R<0.4$ around a muon
 candidate, or any small-$R$ jet is within a region 
 $0.2<\Delta R<0.4$ around an electron candidate, the corresponding electron or muon candidate is discarded.
 Small-$R$ track-jets are defined by applying the same jet reconstruction algorithms to inner-detector tracks
 treated as having the pion mass, and used to avoid overlap between \vvjj{} sideband regions and searches
 for Higgs boson pair production, as discussed in Section~\ref{sec:backgrounds}.

For
the large-$R$ jets, 
the original constituents are calibrated using the local cluster weighting algorithm~\cite{Barillari:1112035} and reclustered using the $k_\perp$
algorithm~\cite{Catani:1993hr} with a radius parameter of $R_\text{sub-jet}=0.2$, to form a
collection of sub-jets. A sub-jet is discarded if it carries less
than $5\%$ of the \pt of the original jet. The
constituents in the remaining sub-jets are then used to recalculate
the large-$R$ jet four-momentum, and the jet energy and mass are further
calibrated to particle level using correction factors derived
from simulation~\cite{Aad:2013gja}. The resulting ``trimmed''\cite{Krohn2010}  large-$R$ jets
are required to have $\pt>200\,\gev$ and $|\eta|<2.0$.
Large-$R$ jets are required to have  an angular separation of $\Delta R>1.0$ from electron candidates.

The large-$R$ jets are used to reconstruct the hadronically decaying
$W/Z$ (``$V$'') boson. A boson
tagger~\cite{ATL-PHYS-PUB-2015-033,CERN-PH-EP-2015-204,Larkoski2014,Larkoski2013} is subsequently used to 
distinguish the  boosted hadronically decaying $V$ boson from jets
originating from quarks (other than the top-quark) or gluons. The tagger is based on the mass
of the jet $m_J$ and a variable $D^{(\beta=1)}_2$, defined in Ref.~\cite{Larkoski2014}, that is sensitive to the
compatibility of the large-$R$ jet with a two-prong decay topology.
The large-$R$ jet is identified by the boson  tagger as a $W$ ($Z$) candidate with its mass within 15\,\gev\  of the
expected $W$ ($Z$) mass peak, which is estimated from simulated events
to be $83.2\gev$ ($93.4\gev$). Large-$R$ jets with mass within 15\,\gev\ from both the $W$ and $Z$ peaks are assigned both hypotheses.
For context, the resolution ranges from $8 \gev$ to $15 \gev$ in the
jet $\pt$ range used in the analysis.
Additionally, a \pt -dependent selection on the \DTwoBetaOne\ variable is configured
so that the average identification efficiency for longitudinally polarised, hadronically decaying $W$ or $Z$  bosons 
is  $50\%$. This selection rejects more 
than $90\%$ of the background.  Large-$R$ track-jets
are defined by applying
the same jet reconstruction and filtering algorithms to inner-detector
tracks treated as having the pion mass.
These jets are ghost-associated to large-$R$ jets and used for the
evaluation of systematic uncertainties, as discussed in
Section~\ref{sec:systematics}.

Small-$R$ jets and small-$R$ track-jets containing $b$-hadrons are identified using the MV2
$b$-tagging algorithm~\cite{BtagMV2c20},  which has an efficiency of 85\%
in simulated $\ttbar$ events. The jets
thus selected are referred to as $b$-jets in the following. The
corresponding misidentification rate for selecting $b$-jet candidates 
originating from a light quark or gluon is less than 1\%. The
misidentification rate for selecting $c$-jets as $b$-jet candidates is
approximately 17\%.

The missing transverse momentum, $\metvec$, with magnitude $\met$, is calculated as the negative vectorial sum of the transverse momenta of 
calibrated objects, such as electrons, muons, and small-$R$ jets. 
Charged-particle tracks compatible with the primary vertex and not matched to any 
of those objects are also included in the \metvec{} reconstruction~\cite{met,met-PUB-2015-023}
. For multi-jet background rejection, a similar quantity, $\mptvec$, is computed using only charged-particle tracks originating from the reconstructed primary vertex to substitute for the calorimeter-based measurements of jet four-momenta. Its magnitude is denoted by $\mpt$.
Both tiers of the ATLAS trigger system also reconstruct \met{}.  The triggers used in this paper reconstruct \met{} based on calorimeter information,
and do not include corrections for muons.

The identification efficiency, energy scale, and resolution of jets, leptons and $b$-jets
are measured in data and correction factors are derived, which are
applied to the simulation to improve the modelling of the data.

\section{Event selection}
\label{sec:selection}

This analysis focuses on identifying diboson events in which at least one vector boson $V$ decays hadronically, and is performed in four different channels identified by the decay of the other vector boson: $\vvjj,\nunuqq,\lnuqq$ and $\llqq$. Event selection criteria are chosen to guarantee the statistical independence of the channels.  The criteria are summarised in Table~\ref{tab:evsel}, and described in more detail below.

Events are selected at trigger level by requiring
at least one large-$R$ jet with $\pt>360\GeV$ in the $\qqqq$ channel, 
large $\met$ in the $\nunuqq$ channel,
large $\met$ or at least one electron in the $\lnuqq$ channel,
and at least one electron or muon in the $\llqq$ channel.
All trigger requirements guarantee full efficiency in the kinematic
region considered in the analysis.
A primary vertex is required to be
reconstructed from at least three charged-particle tracks with
$\pt>400\MeV$.

At least one large-$R$ jet is required, with $\pt>200\GeV$, $|\eta|<2.0$ and $m_J>50\GeV$.
Events are then divided by different pre-selection criteria into
different channels according to the number of ``baseline'' and ``good''  leptons
that are reconstructed.  
A baseline lepton is a loose muon or electron candidate with $\pt>7\GeV$ and $|\eta|<2.7$ 
or $|\eta|<2.47$, respectively, which passes a relaxed set of track-isolation 
and impact parameter requirements.  
A good lepton has $\pt>25\gev$ and is either a muon with $|\eta|<2.5$, or an electron with $|\eta|<2.47$ 
excluding the transition region between barrel and endcap calorimeters ($1.37<|\eta|<1.52$), which passes 
identification and isolation requirements as discussed in Sec. \ref{sec:objects}.

Events with no reconstructed baseline lepton and
with $\met>250\gev$ are assigned to the $\vvqq$ channel. Events
are assigned to the $\vvjj$ channel if they have
no good leptons, $\met<250\gev$, 
 an additional large-$R$ jet
meeting the same selection criteria as the other large-$R$ jet,
 and if the large-$R$ jet with leading $\pT$ satisfies a requirement of
$\pt>450\GeV$ to ensure full trigger efficiency. Events with exactly
one good lepton which satisfies tight track and calorimeter isolation
requirements, and which is either a medium muon or tight electron, or a medium electron with $\pt>300\gev$, are assigned to the $\lvqq$ channel.
Events with
exactly two same-flavour good leptons where one meets medium 
selection criteria, the invariant mass of the dilepton system passes a $Z$ boson mass window selection of $83 < m_{ee}/\GeV < 99$ or $66 < m_{\mu\mu}/\GeV < 116$, 
and, in the case of muons, the two leptons are oppositely charged, are assigned to the $\llqq$ channel. 

Additional event topology requirements are applied to pre-selected events in order to suppress backgrounds.
In the \vvqq\, channel, contributions from non-collision backgrounds and multi-jet production are suppressed by requiring $\MPT>30\gev, |\Delta\phi(\metvec,\mptvec)|<\pi/2$ and by requiring that the minimum azimuthal separation between \metvec{} and any small-$R$ jet is greater than 0.6.

In the \vvjj\, channel, the separation in rapidity between the two
large-$R$ jets,  $|y_{J_1}-y_{J_2}|$, is required to be below  1.2, and their
transverse momentum asymmetry, 
$(\ptJone-\ptJtwo)/(\ptJone+\ptJtwo)$,  is required to be below 0.15.  
To further reduce the multi-jet background, 
large-$R$ jets are required to have $N_\text{trk}<30$ charged-particle tracks with $\pt>500\MeV$,
where the tracks must be consistent with the primary vertex and be matched to the calorimeter jet~\cite{Cacciari:2008gn}.  The matching is made prior to trimming, and is determined by representing each track by a collinear ``ghost'' constituent with negligible energy during jet reconstruction (``ghost association'').

In the \lvqq\, channel, events are required to have no small-$R$ jet identified as a $b$-jet outside a cone of radius $\Delta R=1.0$ around the selected large-$R$ jet to reject backgrounds from \ttbar{} production, and to have $\met>100\gev$ in order to reject multi-jet background. The leptonically decaying $W$ candidate is required to have 
$\ptlv>200\gev$, where the neutrino is assigned transverse momentum $\metvec$ and its momentum along the $z$-axis, $p_z$, is obtained by imposing a $W$ boson mass constraint to the $\ell$--\metvec system.\footnote{The longitudinal momentum $p_z$ is taken to be the smaller in absolute value of the two solutions of the resulting quadratic equation. If a complex value is obtained, the real component is chosen.}
A new resonance with mass $m_{\ell\nu J}$ decaying into two bosons, both
at fairly central rapidity,
would often impart
significant transverse momentum to the bosons relative to the
resonance mass.  The $\pt$ of the two vector-boson candidates is therefore required to have $\ptJ/m_{\ell\nu J}>0.4$ and $\ptlv/m_{\ell\nu J}>0.4$. 
In the \llqq\, channel, similar requirements on the $\pt$ of the two vector-boson candidates 
are applied, namely $\ptJ/m_{\ell\ell J}>0.4$ and  $\ptll/m_{\ell\ell J}>0.4$.

Events are classified as $WW$, $WZ$, or $ZZ$ by applying the corresponding selection criteria to the two boson candidates. If the number of boson-tagged jets exceeds the number of hadronically decaying bosons required by the decay channel, the leading-\pt{} jets are used. 
 The final discrimination between resonant signal
and backgrounds is done in a one-dimensional distribution 
either of mass or of transverse mass. 
In the \vvjj\, channel, the invariant mass
of the jet pair, \mjj, is used in the fiducial region
$1\tev<\mjj<3.5\tev$ whose lower bound is chosen to guarantee full trigger efficiency. In the \vvqq\, channel, the transverse mass of
the $J-\metvec$ system, 
$m_\text{T}=\sqrt{(E_{\text{T},J}+\met)^2-(\mathbf{p}_{\text{T},J}+\metvec)^2}$
is used. In the \lvqq\, channel, $m_{\ell\nu J}$ is used. 
In the \llqq\, channel, the $\pt$ of the dilepton system is scaled event-by-event 
by a single multiplicative factor to set the dilepton invariant mass
$m_{\ell\ell}$ to the mass of the $Z$ boson ($m_Z$) in order to
improve the diboson mass resolution. The invariant mass $m_{\ell\ell
  J}$ is used as the discriminant. 

\begin{figure}[tbp]
 \begin{center}
    \subfigure[]
              {
      \includegraphics[width=0.55\textwidth]{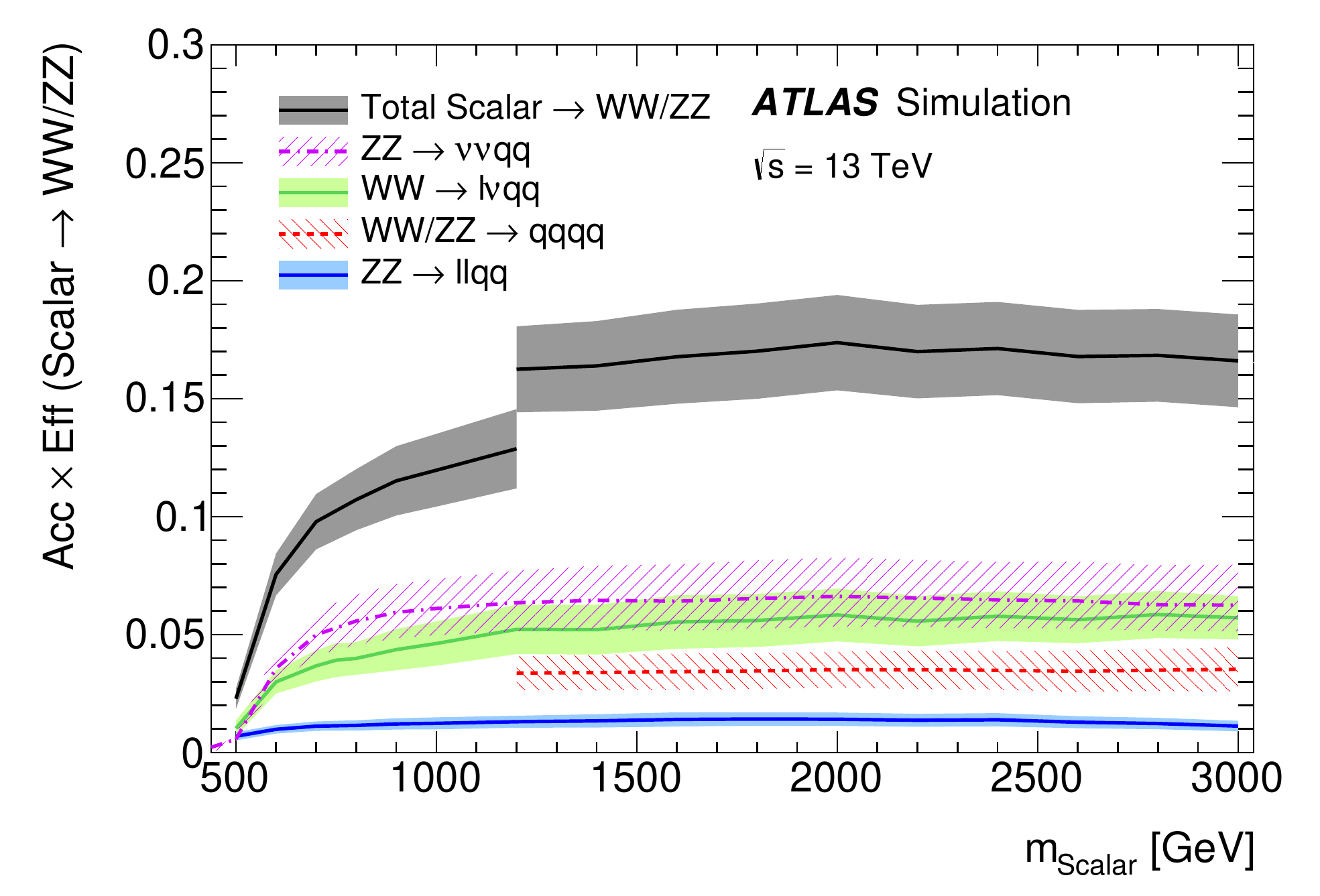}\label{fig:AccEff_ggH_WW_ZZ}
            }
    \subfigure[]
              {
      \includegraphics[width=0.55\textwidth]{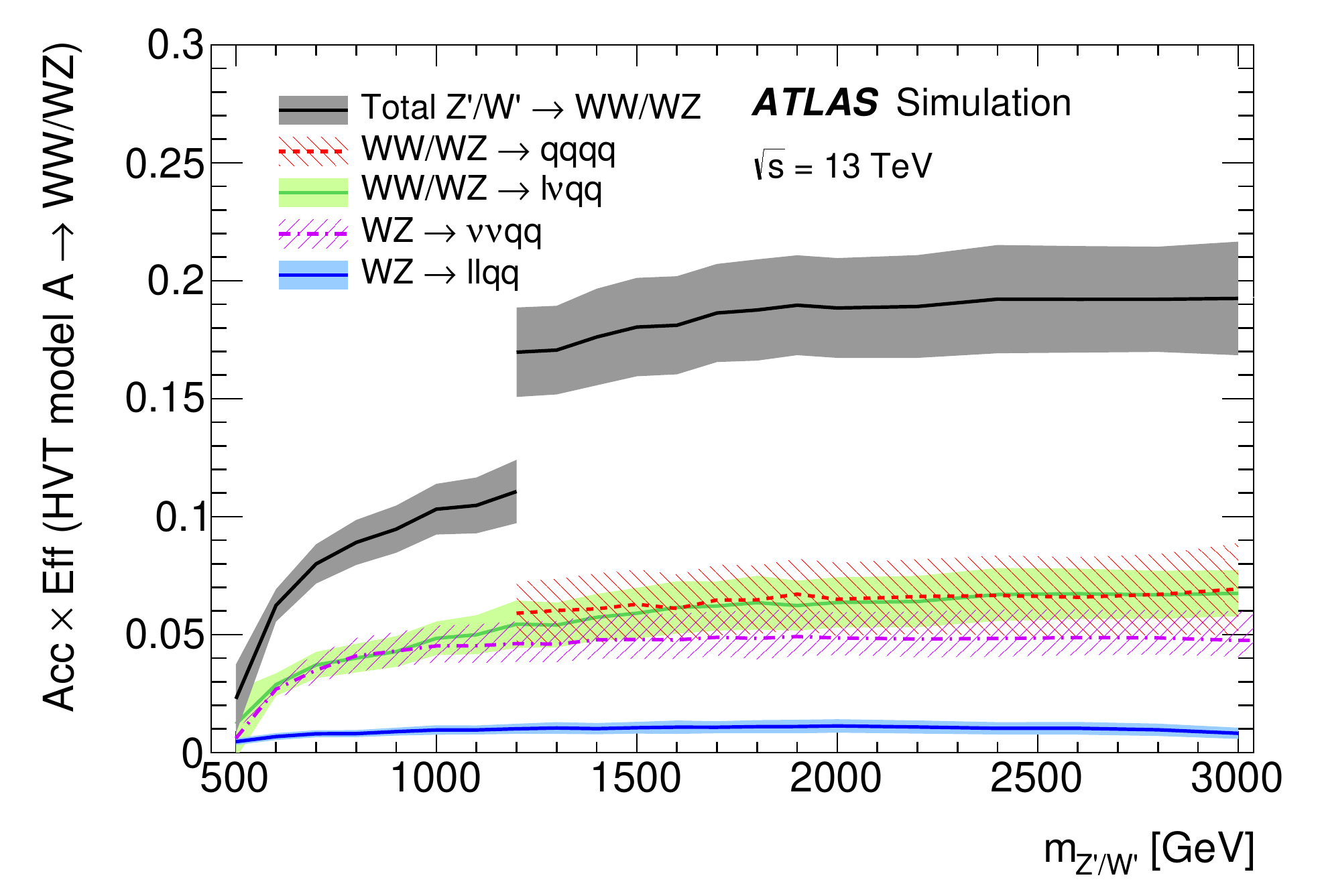}\label{fig:AccEff_HVT_WW_WZ}
              } \\ 
    \subfigure[]
              {
                \includegraphics[width=0.55\textwidth]{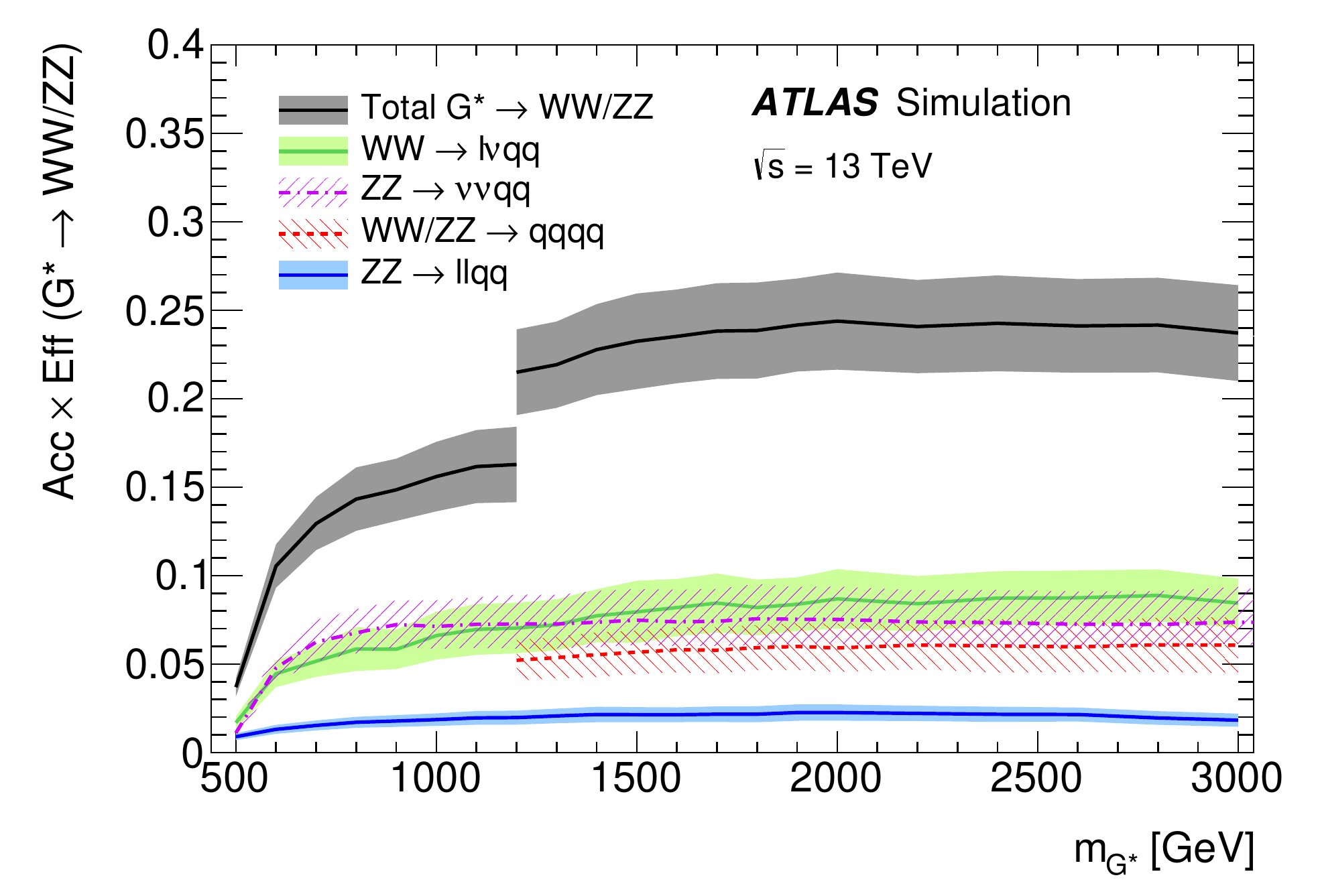}\label{fig:AccEff_RS_WW_ZZ}
    }
  \end{center}
  \vspace{-20 pt}
  \caption{ Signal acceptance times efficiency as a function of the resonance mass, for the different
    channels contributing to the searches for (a) a scalar resonance decaying to $WW$ and $ZZ$, (b) HVT decaying to $WW$ and $WZ$ and (c) bulk RS gravitons decaying to $WW$ and $ZZ$.
    The branching ratio of the new resonance decaying to dibosons
    is included in the denominator of the efficiency calculation. The coloured bands represent 
    the total statistical and systematic uncertainties.
    In the case of the \vvjj{} channel, only signals with resonance masses beyond $1.2\tev$,
    for which the mass peak is fully reconstructed in the fiducial \mjj{} region, are considered.
  }
  \label{fig:acceptance}
\end{figure}

Table \ref{tab:evsel} shows a summary of the event selection criteria in the four channels. 
The combined acceptance times efficiency for a heavy resonance
decaying to dibosons is as large as 18\% for $W'\rightarrow WZ$ and
also for $Z' \rightarrow WW$ in the HVT model-A benchmark assuming
$g_V = 1$.
In the bulk RS benchmark with $\kappa/\bar{M}_\text{Pl} =1$, it
reaches up to 17\% for $G^{*}\rightarrow WW$, and 14\% for
$G^{*}\rightarrow ZZ$. The acceptance times efficiency is estimated with respect to the branching ratio of the signal benchmarks to the 
specific diboson final state and takes into account the $W$ and $Z$ boson branching ratios.
Figure~\ref{fig:acceptance} summarises the acceptance times efficiency
for the different channels as a function of the scalar, HVT, and $G^{*}$ masses,
considering only decays of the resonance into $VV$. The
mass ranges used in the different channels are reflected in the figure.
After all selection criteria are applied, reconstructed diboson mass resolutions for a $W'$ with a mass of 2 TeV, decaying to $WZ$,
are 3\% for \llqq, 5.5\% for \lvqq, and 6\% for \qqqq.

\begin{table}
\begin{center}
\caption{Event selection criteria in the four analysis
  channels. Baseline and good leptons are defined
  in the text.}\label{tab:evsel}
\resizebox{\textwidth}{!}{%
\begin{tabular}{c||c|c|c|c}
\hline
\multirow{2}{*}{Selection level} & \multicolumn{4}{c}{Channel}\\
\cline{2-5}
                     & \vvjj & \vvqq & \lvqq & \llqq \\\hline
\multirow{2}{*}{Trigger}
& Large-R jet,     & $\met$ & $\met (\mu\nu qq)$  & single electron \\
&  $\pt>360\gev$ &        &  or single electron ($e\nu qq$) &  or muon \\\hline
\multirow{3}{*}{Large-R jet}
& $\geq2$, $N_\text{trk}<30$, & \multicolumn{3}{c}{$\geq1$,}  \\
&   $\ptJone>450\gev$,       & \multicolumn{3}{c}{ $\ptJ>200\gev$}\\
&    $\ptJtwo>200\gev$       & \multicolumn{3}{c}{ }\\\hline
 \emph{Baseline leptons} & $0$ & $0$ & $\geq1$ & $\geq2$\\\hline
  \emph{Good leptons} & $0$ & $0$ & $1$ medium $\mu$ or tight$^{\dagger}$ $e$  &
$2$ $e$ or 2 $\mu$, loose + medium\\\hline
\multirow{5}{*}{Topology}
& $\met < 250\gev$,      & $\met > 250\gev$,             & no $b$-jet with $\Delta R(j,J)<1.0$,  & $\ptJ/m_{\ell\ell J}>0.4,$\\
& $\left|y_{J_1}-y_{J_2}\right| < 1.2$,  &  $\mpt > 30\gev$,& $\met>100\gev $,                      &  $\ptll/m_{\ell\ell J}>0.4$, \\
&  $\frac{\ptJone - \ptJtwo}{\ptJone + \ptJtwo}<0.15$  & $|\Delta\phi(\metvec,\MPT)|<\frac{\pi}{2}$,  & $\ptlv>200\gev$,  &$83 < m_{ee}/\GeV < 99$, \\
&                         &  $|\Delta\phi(\metvec,j)|>0.6$ & $\ptJ/m_{\ell\nu J}>0.4$,            & $66 < m_{\mu\mu}/\GeV < 116$\\
&                         &                                & $\ptlv/m_{\ell\nu J}>0.4$       & \\
\hline
Discriminant & \mjj & \mT & $m_{\ell\nu J}$ & $m_{\ell\ell J}$ \\\hline
\end{tabular}
}
\end{center}
\vspace*{-1mm}
{\footnotesize $^\dagger$ The electron, if over 300 \GeV\ in \pT, need only be medium.}
\end{table}

\section{Background estimation}
\label{sec:backgrounds}

The background contamination in the signal regions is different for
each of the channels studied. Different background estimation
strategies are used for the fully hadronic and semileptonic  channels.

In the \vvjj\, channel, the dominant background originates from multi-jet
events, with significantly smaller contributions due to SM
$W/Z+\text{jet}$, diboson, \ttbar\  and single-top-quark production. As all of these processes are expected to produce a smoothly falling \mjj\, spectrum, the overall background is modelled in terms of a probability density function
\begin{equation}
f(x)=N(1-x)^{p_2+\xi p_3}x^{p_3},
\label{eq:qqqq_bkg}
\end{equation}
where $x=\mjj/\sqrt{s}$, $p_2$ and $p_3$ are dimensionless shape parameters, $\xi$ is a constant whose value is chosen to minimise the correlation between $p_2$ and $p_3$, and $N$ is an overall normalisation factor. 
The functional form in Eq. \eqref{eq:qqqq_bkg} is validated using background simulation and validation regions in data, 
defined to be similar to the signal region but with a few differences.
Instead of selecting events where the mass of the large-$R$ jet is consistent with the mass of the $W$ or $Z$ boson, events are selected to have a large-$R$ jet with a mass in the sideband regions, $110-140\gev$ or $50-65\gev$, and without applying the  requirement on the track multiplicity. Specifically, it is required that either both jets have a mass in the range $110-140\gev$ and there are less than two $b$-tagged track-jets matched by ghost-association to either jet, or that one jet has a mass in the range $110-140\gev$ and the other in the range $50-65\gev$. These regions are defined such that the kinematic properties of the selected events are similar to the signal region, and overlap with searches for Higgs boson pair production is avoided.

In the \nunuqq\, channel, the dominant background is \Zjet{} production with
significant contributions from  \Wjet{},
\ttbar, and SM diboson  production.
In the \lvqq\, channel, the dominant backgrounds are \Wjet{} and
\ttbar\ 
production.
In the \llqq\, channel, where two same-flavour leptons with an
invariant mass close to the $Z$ mass are selected, \Zjet{} production
is by far the dominant background.
All three channels also have contributions at the level of a few
percent from single-top-quark and diboson production. 
The single-top-quark process contributes 15\% of the total 
top-quark background in the \lvqq\, channel, 10\% in the the \nunuqq,
and a negligible amount for the \llqq\, channel.
The multi-jet background enters the signal regions of the semileptonic
channels through semileptonic 
hadron decays and through jets misidentified as leptons, and this background is
found to be negligibly small in all  three channels.

In the \nunuqq, \lvqq, and \llqq{} channels, the modelling of $W/Z$+jets backgrounds
is constrained using dedicated control regions.  A region enriched in \Wjet{} events
is used to control the \Wjet{} background normalisation in the \nunuqq{} and \lvqq{} channels;
events in this region are required to fall in the sidebands of the $m_J$ distribution and to
have one reconstructed good muon.  A region enriched in \Zjet{} events is used to control the \Zjet{}
backgrounds in the \nunuqq{} and \llqq{} channels; events in this region are also required to fall
in the sidebands of the $m_J$ distribution, but to have two reconstructed good leptons.

The \ttbar\  background is estimated in the \nunuqq\, and \lvqq\, channels using a control region enriched in top-quark pairs.
This control region is
defined as the \Wjet{} control region, without the $m_J$ sideband criterion and with the added requirement of at least one additional $b$-jet
with a separation $\Delta R>1$ from the large-$R$ jet.
The \ttbar\ background for the \llqq\, channel is estimated from MC simulation.

The $W$, $Z$ and \ttbar{} control regions are included in the combined profile
likelihood fit described in Section \ref{sec:combination} to help constrain the $W$+jets, $Z$+jets and \ttbar\  normalisation in the signal regions.

The diboson contributions to the \nunuqq, \lvqq\, and \llqq\, channels are estimated using MC simulation. Single-top-quark production is constrained by the $\ttbar$ control region using the same normalisation factor as for $tt$.

\section{Systematic uncertainties}
\label{sec:systematics}

The most important sources of systematic uncertainty are those related to the energy scale and resolution
of the large-$R$ jet \pt, mass, and \DTwoBetaOne. The systematic uncertainties related to the scales of the
large-$R$ jet \pt, mass and \DTwoBetaOne are extracted following the technique described in Ref.~\cite{Aad:2013gja}.
Track-jets are geometrically matched to calorimeter jets, and for each observable of interest, e.g. \pt,
mass, or \DTwoBetaOne, a systematic uncertainty is estimated  from the comparison of the ratio of the
matched track-jet observable to the calorimeter-jet observable between simulation and data. For the jet
\pt{} and mass, $\sqrt{s}=13\tev$ data and simulation are used. For \DTwoBetaOne, $\sqrt{s}=8\tev$
simulation and data are used, and an additional uncertainty is added to account for the differences
between 8 TeV and 13 TeV~\cite{ATL-PHYS-PUB-2015-033}. The  uncertainties in the large-$R$ jet \pt,
mass, and \DTwoBetaOne scale are 5\%, 6\% and 10\%, respectively.  

The resolution of each of these large-$R$ jet observables is determined as the standard deviation of a Gaussian
fit to the distribution of the observable response defined as the
ratio of the calorimeter-jet
observable to  a simulated-particle-level jet observable. The relative uncertainties in these resolutions
are estimated based on previous studies with $\sqrt{s} = 7\tev$ data and $\sqrt{s}=13\tev$ simulation.
For the large-$R$ jet \pt{}~\cite{Aad:2013gja} and mass resolution a 20\% uncertainty is assigned, while for the \DTwoBetaOne{}
resolution a 10\% uncertainty is assigned. The large-$R$ jet mass resolution uncertainty is estimated from 
variations in data and simulation in the widths of the $W$-jet mass peaks in $tt$ events~\cite{Aad:2013gja}.
The \DTwoBetaOne{} resolution uncertainty is estimated by comparing 13 TeV simulation samples from
different generators and shower simulations~\cite{ATL-PHYS-PUB-2015-033}.

Other subdominant experimental systematic uncertainties include those in the lepton energy and momentum scales, in lepton
identification efficiency, in the efficiency of the trigger requirements, and in
the small-$R$ jet energy scale and resolution.  All experimental systematic uncertainties are treated as fully correlated among all channels.

Uncertainties are also taken into account for possible differences between data and the simulation model that is used for each process.  

In the \vvqq\ channel,  an uncertainty on the shape of the \mT{} spectrum for the \Wjet{} and \Zjet{} backgrounds is extracted by comparing the nominal shape obtained with \textsc{Sherpa} to the one obtained with an alternative sample generated with \textsc{MadGraph5\_aMC@NLO}.

In the \lvqq\ channel,  an uncertainty on the shape of the $m_{\ell\nu J}$ distribution of the dominant \Wjet{} background is obtained by comparing the $m_{\ell\nu J}$ shape in  simulation and in data in the \Wjet{} control region after the expected
\ttbar\ and diboson contributions are subtracted. The ratio of the data distribution to that predicted by MC is fitted with a first-order polynomial  and its deviation from unity is used as a modelling uncertainty.

In the \llqq\ channel,  an uncertainty on the shape of the $m_{\ell\ell J}$ spectrum for the \Zjet{} background is assessed by comparing the shape difference between the 
  \textsc{Sherpa} predictions and the data-driven estimate using events in the $Z$ control region.

The data and simulation agree very well for events in the
top-quark control region. The uncertainty in the shape of the mass
distributions for the \vvqq,~\lvqq~and \llqq~channels from the \ttbar\ background is estimated by comparing a sample generated using
a\textsc{MC@NLO}~\cite{Alwall:2014hca} interfaced with \textsc{Pythia} 8.186 to the nominal \ttbar\  sample. 
Additional systematic uncertainties in parton showering and hadronisation are evaluated by comparing the
nominal sample showered with \textsc{Pythia} to one showered with {\sc Herwig}~\cite{Bahr:2008pv}.
Samples of \ttbar{} events generated with the factorisation and
renormalisation scales doubled or halved are compared to the nominal
sample, and the largest difference observed in the mass discriminants
is taken as an additional uncertainty arising from the QCD scale uncertainty.

Theoretical uncertainties in the SM diboson production cross-section, including the effect of PDF and scale uncertainties, are taken into account and amount to about 10\%~\cite{Campbell:1999ah}.
An  uncertainty in the shape of the predicted diboson $m_{\ell\ell J}$
spectrum in the \llqq{} channel is derived by comparing MC samples
generated by \textsc{Sherpa} and \textsc{POWHEG BOX}. Shape uncertainties are found to have negligible impact in the \vvqq{} and \lvqq{} channels. 

The uncertainties in the modelling of the \Zjet{} and \Wjet{}
backgrounds are treated as uncorrelated since they are evaluated
differently in each channel. For the $tt$ background, the modelling uncertainty is treated as correlated between the \lvqq{} and \llqq{} channels, and uncorrelated with the modelling uncertainty in the \vvqq{} channel. The diboson normalisation uncertainty is taken as correlated among the \vvqq{}, \lvqq{} and \llqq{} channels.

Uncertainties in the signal acceptance arise from the choice of PDF
and from the amount of initial- and final-state radiation present in
simulated signal events.  The PDF-induced  uncertainties in the signal
acceptance for semileptonic decay channels are derived using the
PDF4LHC recommendations~\cite{Botje:2011sn}; in all channels the
resulting uncertainty is at most 4\%.  PDF-induced  uncertainties are not evaluated for the \qqqq{} channel, where they are subdominant to other acceptance effects.  The uncertainty in the integrated luminosity has an impact of $5\%$ on the signal normalisation.  All signal acceptance uncertainties are treated as fully correlated across all search channels.  

The  uncertainty in modelling  background distribution shapes  in the \vvjj~channel is found to be negligible compared to statistical uncertainties in the background fit parameters. 
An additional
uncertainty in the signal normalisation is
introduced in the \vvjj{} channel to take into account potentially
different efficiencies of the $N_\text{trk}<30$ requirement in data and
simulation.
This uncertainty is estimated in a data control sample enriched in
$W/Z$+jets events, where the $W/Z$ bosons decay to quarks. This
control sample is obtained by applying the \DTwoBetaOne{} selection
only to the highest-\pt{} large-$R$ jet in dijet events. The $m_J$
distribution is fitted in subsamples with different track multiplicity
selections to obtain the rates of  $W/Z$ decays in each
sample. From these the uncertainty from the track
multiplicity cut 
is estimated to be 6\%. 

For all the considered signal hypotheses, the impact of each source of uncertainty on the search is evaluated in terms of the corresponding contribution to the total uncertainty in the fitted number of signal events, as obtained after the statistical procedure described in the next section. The dominant contribution is due to large-$R$ jet scale uncertainties and amounts to about $35\%$ of the total uncertainty. Additional contributions are due to uncertainties in the modelling and normalisation of backgrounds in the \vvqq{}, \lvqq{} and \llqq{} channels (about $20\%$), and small-$R$ jet energy scale uncertainties (about $10\%$). Sub-leading contributions have an overall impact of less than about $15\%$.

\section{Statistical analysis}
\label{sec:combination}

 In the combined analysis to search for a scalar resonance decaying to
 $WW$ or $ZZ$, HVT decaying to $WW$ or $WZ$, and bulk \grav{} decaying
 to $WW$ or  $ZZ$, all four individual channels  
 are used. Table~\ref{tab:gravandwpparticipants} summarises the signal region and mass range in which the individual channels contribute to the search. 
 
 \begin{table}[!pht]
\caption{
  Channels, signal regions and mass ranges where the channels contribute to the search.
\label{tab:gravandwpparticipants}
}
\begin{center}
\begin{tabular}{c|c|c|c|c}
\hline
\hline
\multirow{2}{*}{Channel} & {Signal region} &   {Scalar} & {HVT $W'$ and $Z'$} & {\grav{}} \\
 &(selection)  &   mass range [\TeV] & mass range [\TeV] &   mass range [\TeV] \\
\hline
\multirow{2}{*}{$qqqq$} & $WW+ZZ$ & 1.2--3.0 & --  &  1.2--3.0 \\
& $WW+WZ$ & -- & 1.2--3.0 & -- \\
\hline
\multirow{2}{*}{$\nu \nu qq$} & $WZ$ & -- & 0.5--3.0 & -- \\
& $ZZ$& 0.5--3.0 & -- &  0.5--3.0  \\
\hline
\multirow{2}{*}{\lnuqq} & $WW+WZ$ & -- & 0.5--3.0  & --   \\
& $WW$ & 0.5--3.0 & -- &  0.5--3.0  \\
\hline
\multirow{2}{*}{\llqq}  & $WZ$ &  -- & 0.5--3.0 & -- \\
& $ZZ$  & 0.5--3.0 & -- & 0.5--3.0 \\
\hline
\hline
\end{tabular}
\end{center}
\end{table}

The statistical interpretation of these results uses the data modelling and handling toolkits 
RooFit~\cite{Verkerke:2003ir}, RooStats~\cite{Moneta:2010pm} and HistFactory~\cite{Cranmer:2012sba}.
It proceeds by
defining the likelihood function $L(\mu, \vec\theta)$ for a
particular model, with an implicit   signal description,   in
terms of the signal  strength $\mu$, and the additional  set of nuisance
parameters $\vec\theta$ which can be related to both background and signal.
The likelihood function is computed considering in each channel bins of the discriminating variable; the binning is chosen based on the expected mass resolution and statistical uncertainty, as estimated from simulation.
The nuisance
parameters are either free to float, as in the case of the $p_2$ and
$p_3$ parameters 
used in the $qqqq$ channel to estimate the background, or  
constrained from external studies represented by Gaussian terms.
The likelihood for the
combination of the four search channels is the product of the Poisson likelihoods for the
individual channels, except in the case of common nuisance parameters,

\begin{equation}
  L(\mu,\vec\theta) = \prod_c \prod_i \mbox{Pois} \left( n^{\text{obs}}_{c_i} |
  n^{\text{sig}}_{c_i}(\mu,\vec{\theta}) +
  n^{\text{bkg}}_{c_i}(\vec{\theta})   \right)
    \prod_k f_k(\theta'_k | \theta_k).
  \end{equation}

The terms $n^{\text{obs}}_{c_i}$ represent the number of events observed, and the terms $n^{\text{sig}}_{c_i}$,  $n^{\text{bkg}}_{c_i}$, 
the number of events expected from signal or background in bin $i$ of the discriminant from channel $c$. The term
$f_k(\theta'_k | \theta_k$ ) represents the set of constraints on $\vec\theta$ from auxiliary measurements $\theta'_k$:
these constraints include normalisation and shape uncertainties in the
signal and background models, and, except for the \qqqq\ channel,
include the statistical uncertainties of the simulated bin content.
The \Wjet{} normalisation is a free parameter in the combined likelihood fit in all the channels. The normalisation of the \Zjet{} background in the \llqq{} and \vvqq{} channels is a free parameter in the combined likelihood fit. In the \lvqq{} channel, where the contribution from \Zjet{} is small, the normalisation obtained from MC simulation is used instead, with an 11\% systematic uncertainty assigned.
The $tt$ normalisation in the \lvqq{} and \vvqq{} channels is a free parameter in the combined likelihood fit. In the \llqq{} channel, where the \ttbar{} background contribution is small, its normalisation is 
based on the theoretical cross-section with a 10\% systematic uncertainty assigned.

The likelihood function $L(\mu,\vec{\theta})$ is used to construct the
profile-likelihood-ratio  test statistic~\cite{ATL-PHYS-PUB-2011-011}, defined as:

\begin{equation}
  t = -2 \ln   \lambda(\mu) = -2 \ln \left( \frac{L\big(\mu\,,\,\hat{\hat{\vec{\theta}}}(\mu)\big)}
  {L(\muhat,\hat{\vec{\theta}})}
                              \right) \label{eq:LH},
\end{equation}

where $\muhat$ and $\hat{\vec{\theta}}$ are the values of the
parameters that maximise the likelihood function $L(\mu,
\vec{\theta})$ globally, and  $\hat{\hat{\vec{\theta}}}(\mu)$ are the
values of $\vec{\theta}$ which maximise the likelihood function given a
certain value of $\mu$. 
The parameter $\muhat$ is required to be non-negative. 
This test statistic is used to derive the statistical results of the analysis.

For calculating $p$-values, which test the compatibility of the data with the
background-only model,  the numerator of Eq.~(\ref{eq:LH})
is evaluated for the background-only hypothesis, i.e. signal strength $\mu=0$.
In extracting upper limits,  the calculation is modified such
that if $\muhat>\mu$, $\lambda(\mu)$ is taken to be 1; this ensures
that a signal larger than expected  is not taken as evidence against a
model. The asymptotic distributions of the above test statistic are
known and described in Ref.~\cite{Asymptotic}, and this methodology is
used to obtain the results in this paper.

Upper limits on the production cross-section times branching ratio
to diboson final states for the benchmark signals are set using the
modified-frequentist $CL_s$ prescription~\cite{cls}, where the
probability of observing  $\lambda$ to be larger than a particular
value, is calculated 
using a one-sided profile likelihood. 
The calculations are done using the lowest-order asymptotic
approximation, which was validated to 
better than 10\% accuracy using pseudo-experiments. All
limits  are set at the  95\% confidence level (CL).

\section{Results}
\label{sec:results}

\begin{figure}[th!]
 \begin{center}
    \subfigure[]
              {
      \includegraphics[width=0.45\textwidth]{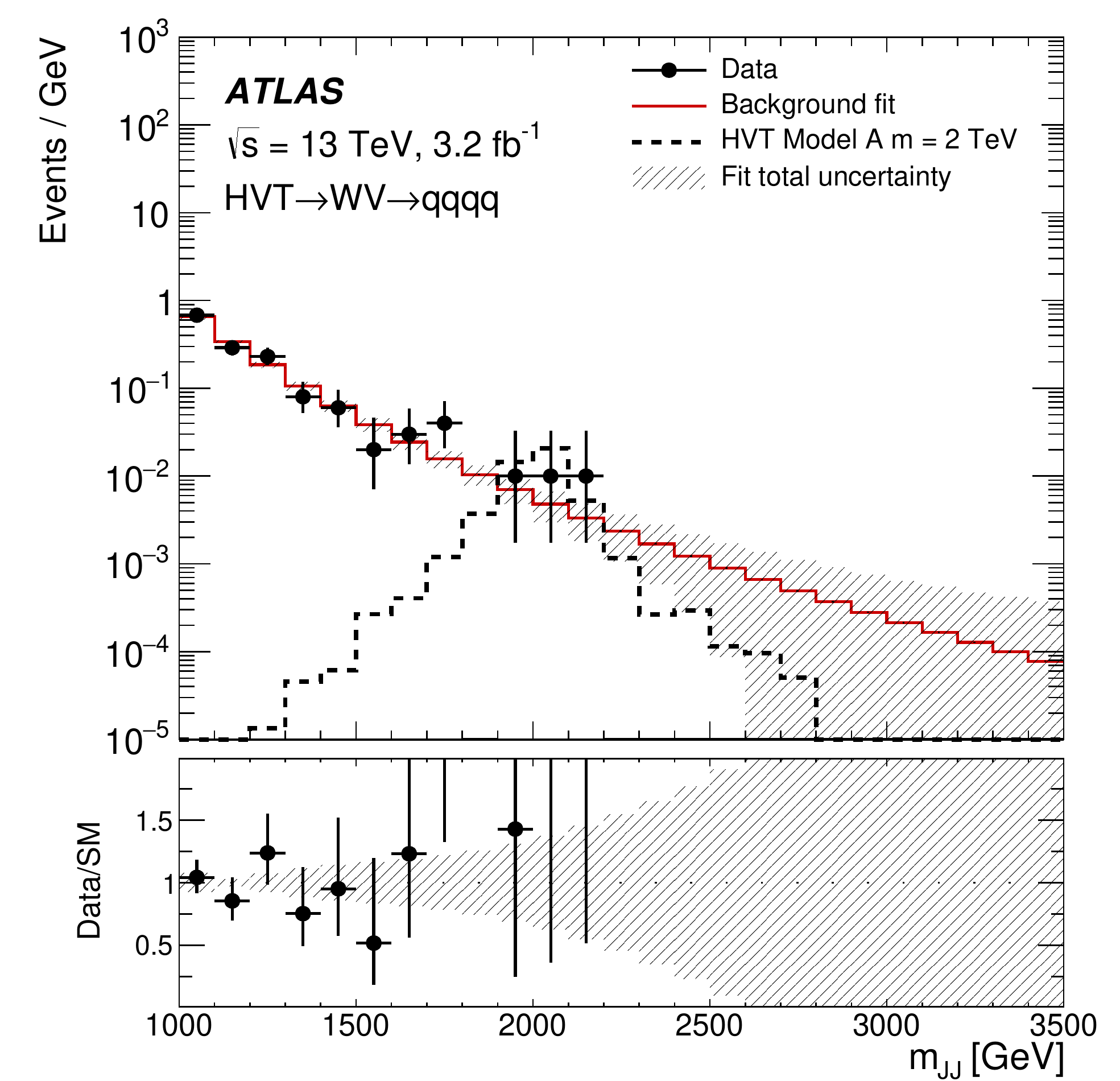}\label{fig:qqqq_postFit_HVT}
              }
    \subfigure[]
              {
                \includegraphics[width=0.45\textwidth]{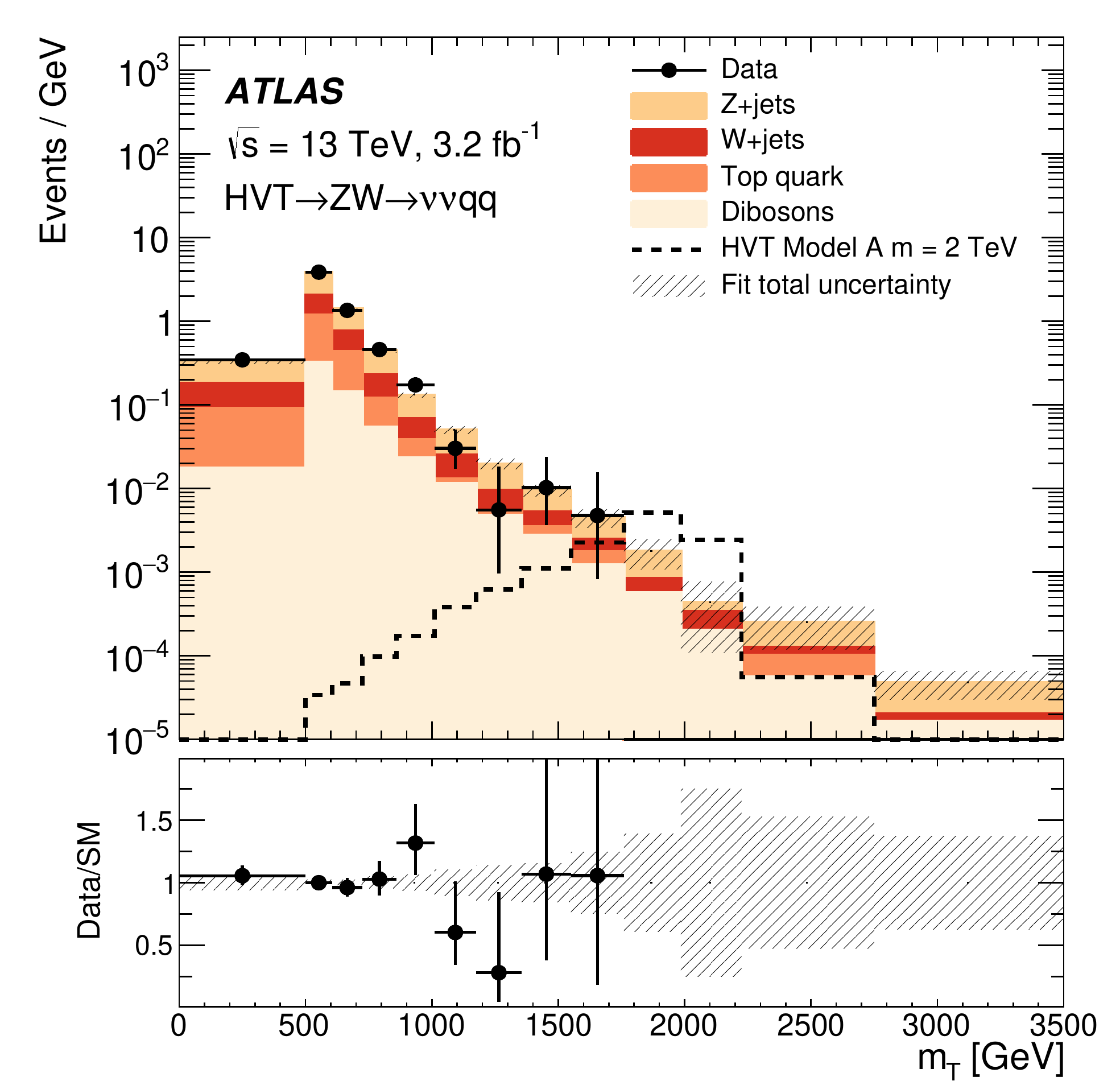}\label{fig:vvqq_postFit_HVT}
    } \\ 
    \subfigure[]
              {
      \includegraphics[width=0.45\textwidth]{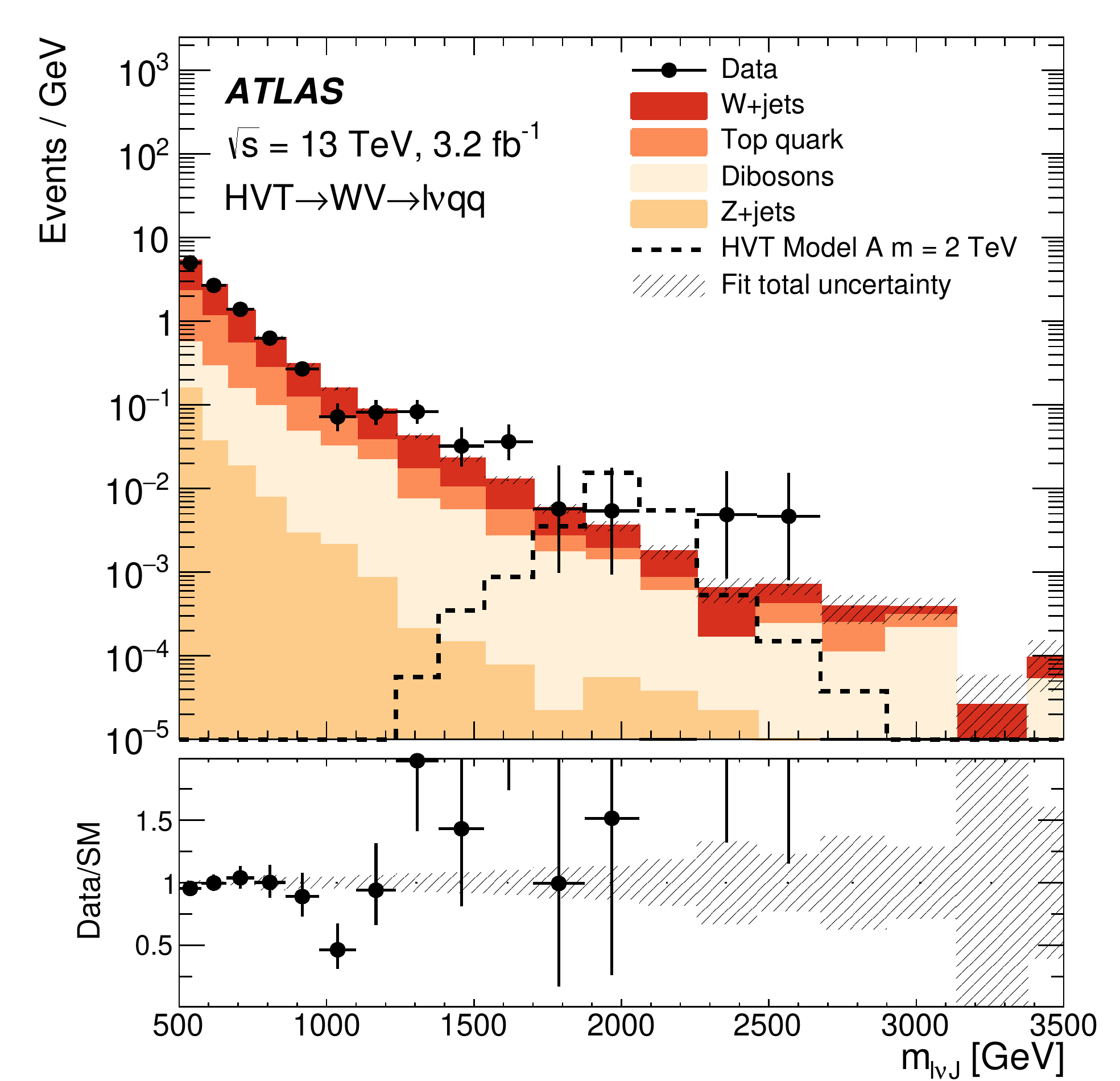}\label{fig:lvqq_postFit_HVT}
              }
    \subfigure[]
              {
                \includegraphics[width=0.45\textwidth]{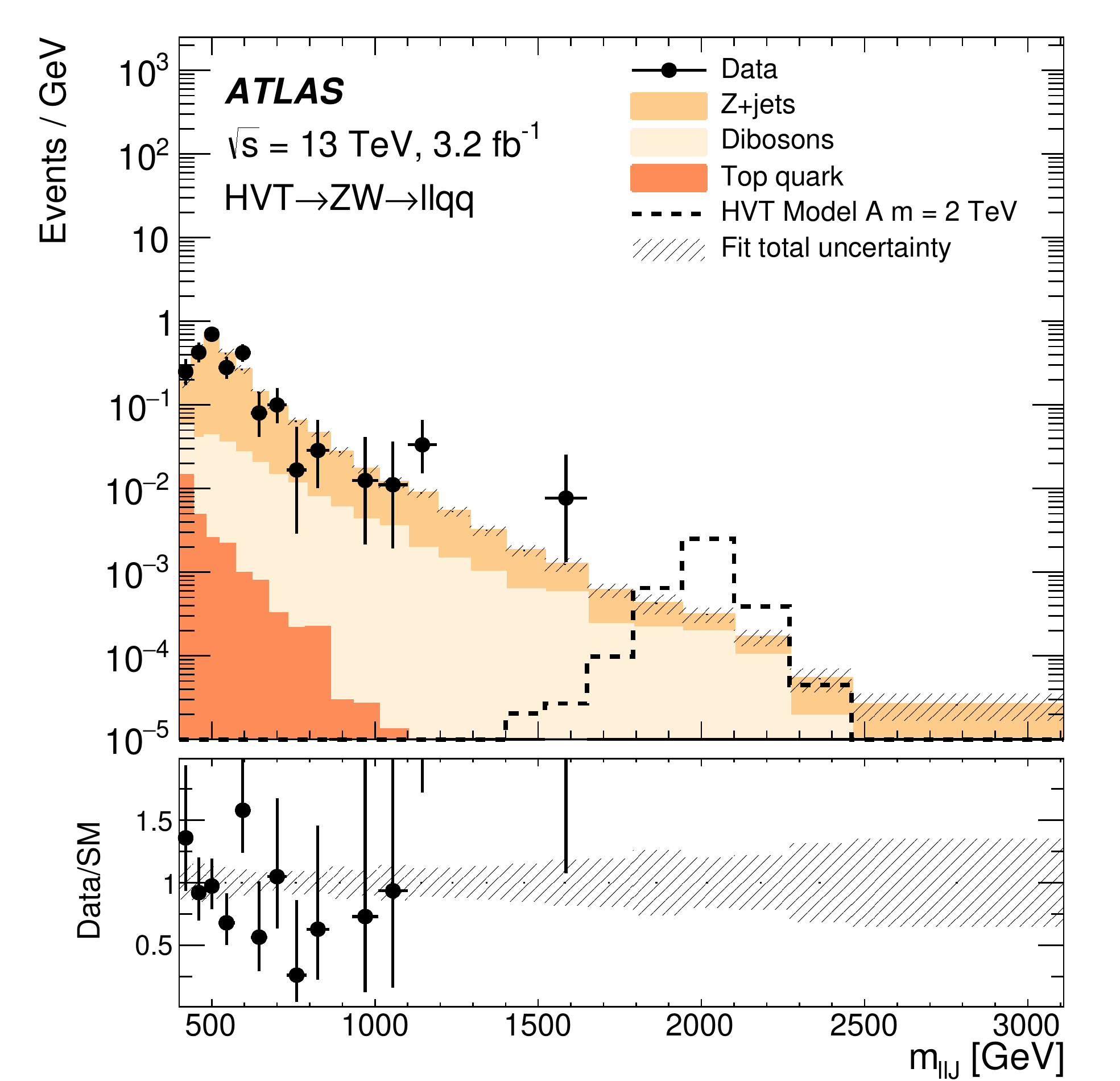}\label{fig:llqq_postFit_HVT}
    } 
  \end{center}
  \vspace{-20 pt}
  \caption{Distribution of the data compared to the background
    estimate for the analysis discriminant in the signal regions for
    the HVT search; (a) the $m_{JJ}$ distribution in the $qqqq$
    channel, (b) the $m_{\mathrm T}$ distribution in the $\nu \nu qq$ channel,
    (c) $m_{\ell \nu J}$ in the $\ell \nu qq$ channel, and (d)
    $m_{\ell \ell J}$ in the $\ell \ell qq$ channel.  The ``Top
    quark'' distribution includes both the \ttbar{} and single-top-quark
    backgrounds.  The upper panels show the distribution of the
    observed data and estimated backgrounds as a function of the
    analysis discriminants. The observed data are shown as points,
    solid colours represent the different background contributions and
    the shaded bands reflect the systematic uncertainties in the
    estimated background. The lower panels show the ratio of the
    observed data to the estimated background as a function of the
    analysis discriminant.  The decay modes ``$WV$'' or ``$ZW$''
    indicate the mass requirements placed on the hadronically decaying
    boson, where a ``$W$'' or ``$Z$'' indicates a narrow mass window
    around the corresponding boson mass, and a ``$V$'' indicates the
    wider mass window including both the $W$ and $Z$ boson masses.
    More details are given in the text.
  }
  \label{fig:postFit_HVT}
\end{figure}

\begin{figure}[th!]
  \begin{center}
    \subfigure[]
              {
      \includegraphics[width=0.45\textwidth]{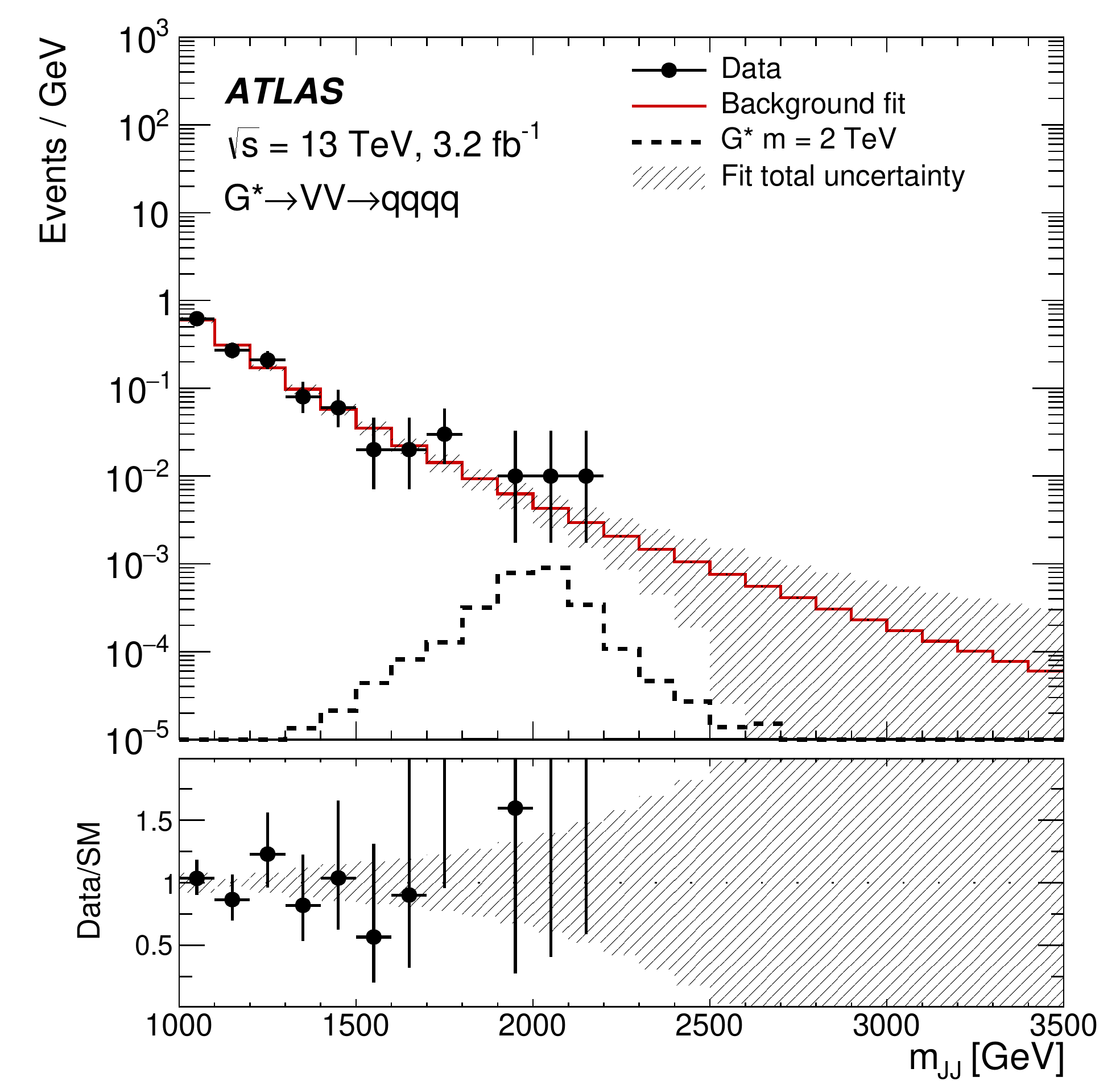}\label{fig:qqqq_postFit_Grav}
              }
    \subfigure[]
              {
                \includegraphics[width=0.45\textwidth]{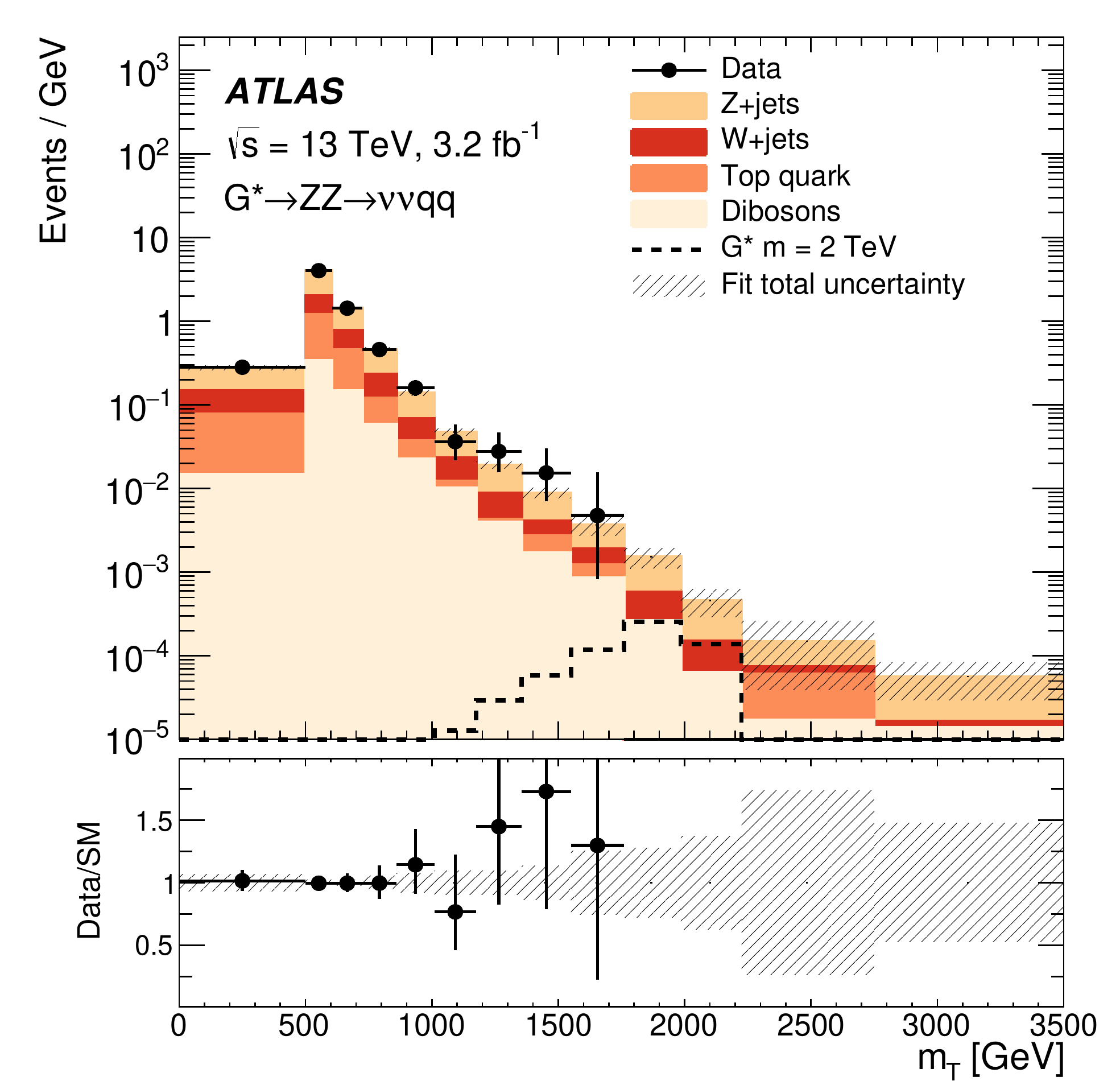}\label{fig:vvqq_postFit_Grav}
    } \\ 
    \subfigure[]
              {
      \includegraphics[width=0.45\textwidth]{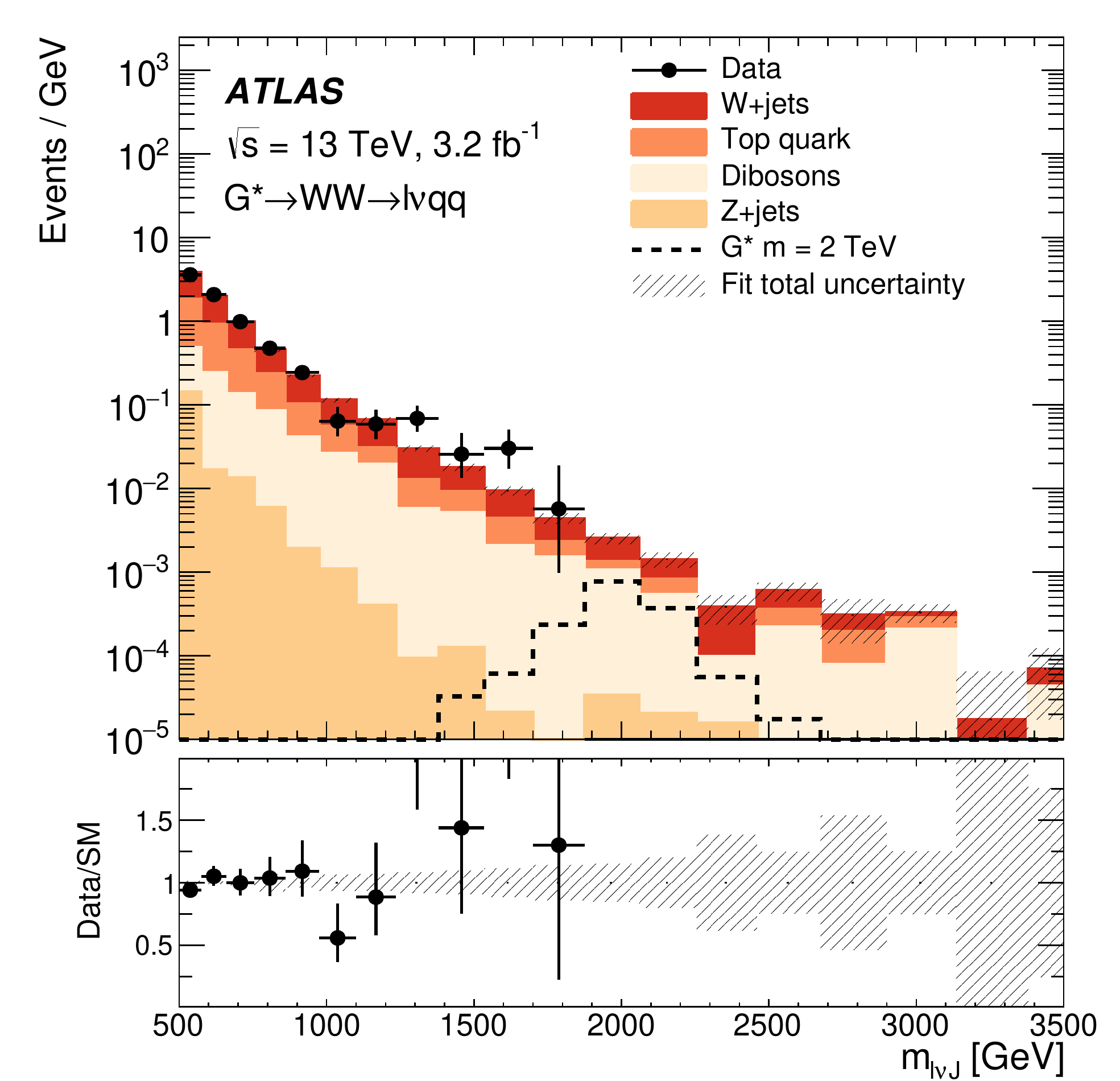}\label{fig:lvqq_postFit_Grav}
              }
    \subfigure[]
              {
                \includegraphics[width=0.45\textwidth]{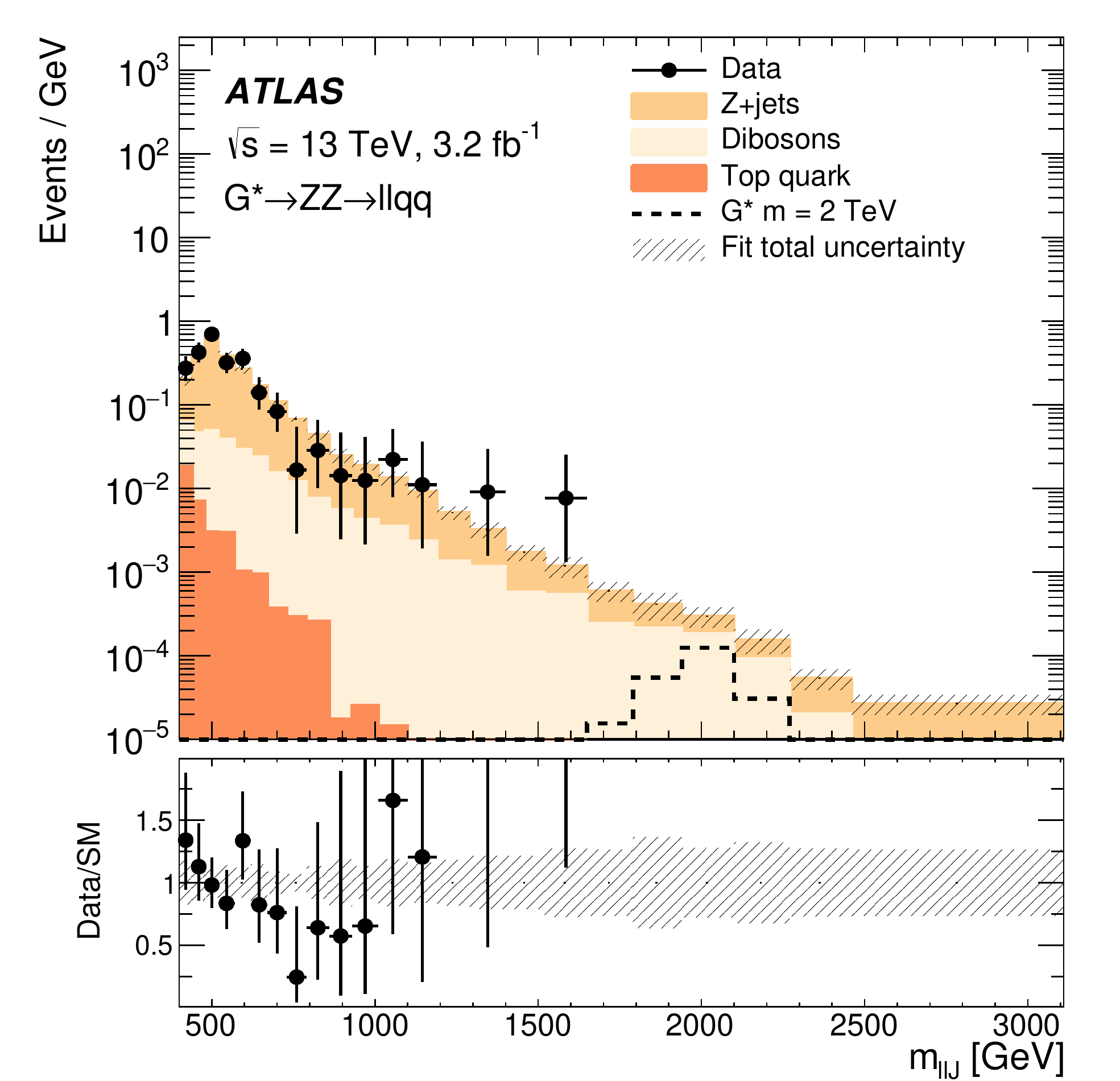}\label{fig:llqq_postFit_Grav}
    } 
  \end{center}
  \vspace{-20 pt}
  \caption{Distribution of the data compared to the background
    estimate for the analysis discriminant in the signal regions for
    the scalar and bulk RS \grav{} searches; (a) the $m_{JJ}$
    distribution in the $qqqq$ channel, (b) the $m_{\mathrm T}$ distribution in
    the $\nu \nu qq$ channel, (c)  $m_{\ell \nu J}$ in the $\ell
    \nu qq$ channel, and (d) $m_{\ell \ell J}$ in the $\ell \ell qq$
    channel.  The ``Top quark'' distribution includes both the \ttbar{}
    and single-top-quark backgrounds.  The upper panels show the
    distribution of the observed data and estimated backgrounds as a
    function of the analysis discriminants. The observed data are
    shown as points, solid colours represent the different background
    contributions and the shaded bands reflect the systematic
    uncertainties in the estimated background. The lower panels show
    the ratio of the observed data to the estimated background as a
    function of the analysis discriminant.  The decay modes ``$WW$'',
    ``ZZ'', or ``$VV$'' indicate the mass requirements placed on the
    hadronically decaying boson, where a ``$W$'' or ``$Z$'' indicates
    a narrow mass window around the corresponding boson mass, and a
    ``$V$'' indicates the wider mass window including both the $W$ and
    $Z$ boson masses.  More details are given in the text.
  }
  \label{fig:postFit_Grav}
\end{figure}

The background estimation techniques described in
Section~\ref{sec:backgrounds} are applied to the selected data, and
the results in the four different analysis channels are shown in
Figure~\ref{fig:postFit_HVT} for the channels relevant to the HVT
($WZ,WW$) search and in Figure~\ref{fig:postFit_Grav} for those
relevant to  the scalar resonance and bulk RS $\grav \to WW,ZZ$ searches, respectively.  Both figures represent background-only fits to the data.
The total yields in the different signal and control regions for the HVT $WZ$ channel are also shown in Table~\ref{tab:yields_WZ}.  
Good agreement is found between the data and the background-only
hypothesis. The most significant excess over the expected background is
observed in the \largestPzeroSel{} selection for a mass of
$1.6\tev{}$, with a $p$-value equivalent to a local
significance of \largestPzeroVal{} standard deviations.
Upper limits at the 95\% CL are set on the production cross-section
times the branching ratio of new resonances decaying to diboson final
states.

\begin{table}[tbp]
\renewcommand{\arraystretch}{1.25}
\caption{Expected and observed yields in signal and control regions
  for the $W'\rightarrow WZ$ signal hypothesis.  Yields and
  uncertainties are evaluated after a background-only fit to the data.
  The background for the $qqqq$ channel is evaluated \textit{in situ}
  and only the total background yield is indicated.  The $W$+jets
  background for the $Z$+jets control region and the $\ell\ell qq$
  signal region is negligible.  The uncertainty in the total
  background estimate can be smaller than the  sum in quadrature of the individual background contributions due to anti-correlations between the estimates of different background sources.}
\resizebox{\textwidth}{!}{%
\begin{tabular}{ c | c@{\ $\pm$\ }c c@{\ $\pm$\ }c c@{\ $\pm$\ }c | c@{\ $\pm$\ }c  c@{\ $\pm$\ }c c@{\ $\pm$\ }c c@{\ $\pm$\ }c }
\hline
\hline
                      &\multicolumn{6}{c|}{Control Regions}  &\multicolumn{8}{c}{Signal Regions}\\
\cline{1-15}
                      &\multicolumn{2}{c}{         $W$+jets}&\multicolumn{2}{c}{         $Z$+jets}&\multicolumn{2}{c|}{         $\ttbar$}&\multicolumn{2}{c}{      $\nu\nu qq$}&\multicolumn{2}{c}{    $\ell\ell qq$}&\multicolumn{2}{c}{     $\ell\nu qq$}&\multicolumn{2}{c}{           $qqqq$}\\
\hline
            SM Diboson&                     53 &           8&                     15 &           4&                     12 &           3&                     70 &           8&                     12 &           2&                     64 &           9&\multicolumn{2}{c}{     }             \\
  $\ttbar$, single-$t$&                    325 &          49&                    1.4 &         0.8&                    780 &          34&                    170 &          15&                    1.2 &         0.9&                    230 &          31&\multicolumn{2}{c}{     }             \\
              $Z$+jets&                     17 &           3&                    387 &          19&                    2.5 &         0.7&                    385 &          24&                    102 &           7&                     11 &           2&\multicolumn{2}{c}{     }             \\
              $W$+jets&                    797 &          66&\multicolumn{2}{c}{     }             &                     54 &          13&                    208 &          23&\multicolumn{2}{c}{     }             &                    397 &          31&\multicolumn{2}{c}{     }             \\
\hline
      Total Background&                   1193 &          31&                    403 &          19&                    849 &          29&                    832 &          26&                    115 &           7&                    702 &          20&                    128 &          11\\
\hline
              Observed&\multicolumn{2}{c}{1200}             &\multicolumn{2}{c}{ 406}             &\multicolumn{2}{c|}{ 848}             &\multicolumn{2}{c}{ 838}             &\multicolumn{2}{c}{ 109}             &\multicolumn{2}{c}{ 691}             &\multicolumn{2}{c}{ 128}             \\
\hline
\hline
\end{tabular}
}
\label{tab:yields_WZ}
\end{table}

\begin{figure}[th!]
  \begin{center}
       \includegraphics[width=0.6\textwidth]{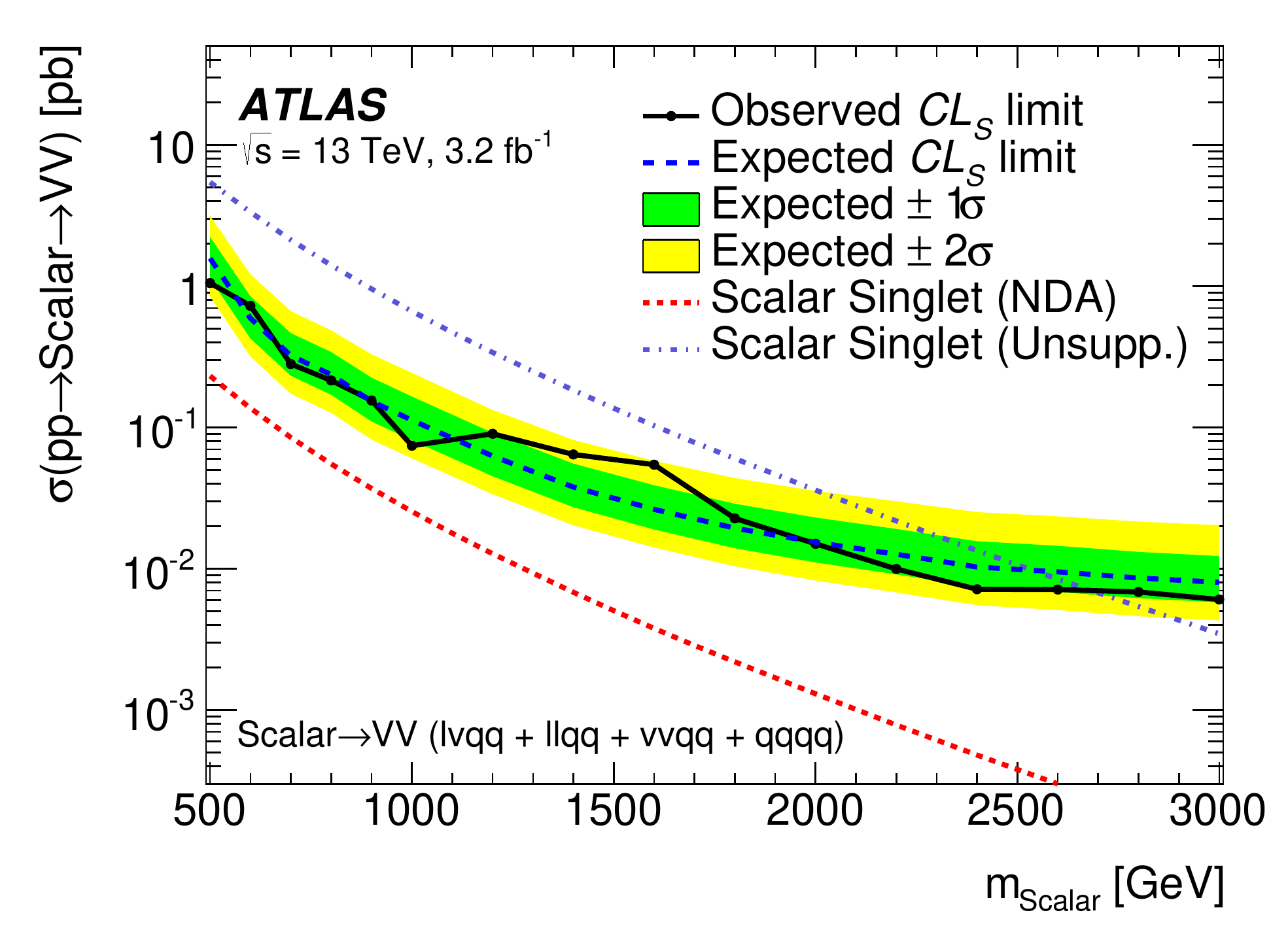}\label{fig:scalar_limit}
  \end{center}
  \vspace{-20 pt}
  \caption{ Observed and expected $95\%$ CL limits on the cross-section times branching ratio to diboson final states for a narrow-width
    scalar resonance, as a function of its mass, combining the $WW$ and
      $ZZ$ decay modes.
  }
  \label{fig:scalarLims}
\end{figure}

\begin{figure}[th!]
  \begin{center}
      \includegraphics[width=0.6\textwidth]{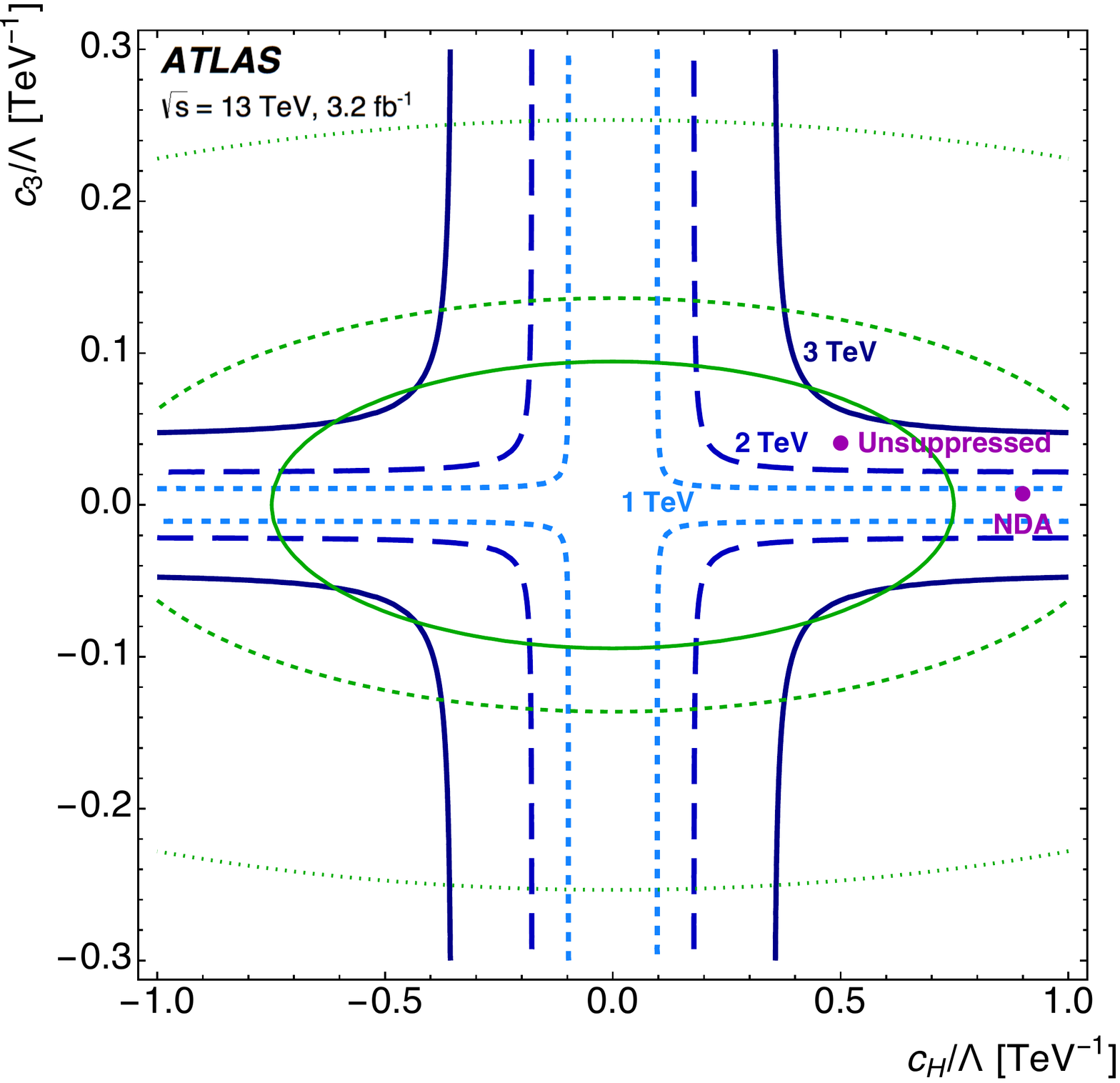}
  \end{center}
  \vspace{-20 pt}
  \caption{Observed 95\% CL exclusion contours in the parameter space
    $(c_H/\Lambda,c_3/\Lambda)$ for scalar resonances of mass $1\tev$, $2\tev$ and
    $3\tev$. The  region inside each green  ellipse indicates, for the
    corresponding assumed mass, the part of the parameter space in
    which the ratio of the resonance's  total width $\Gamma$ to its mass
    $m$ is below $5\%$, which is comparable to the experimental
    mass resolution. Points inside the ellipse, but where the absolute values of the $c_H/\Lambda$ and $c_3/\Lambda$ parameters are larger than at the exclusion contour, are considered to be excluded at a CL greater than 95\%. 
  The solid lines correspond to results for a resonance mass of
  3~\TeV; the long-dashed lines correspond to a resonance mass of
  2~\TeV; the short-dashed lines correspond to a resonance mass of
  1~\TeV. Parameters for the Unsuppressed and NDA  benchmark models are also shown.}
  \label{fig:scalarparams}
\end{figure}

Figure~\ref{fig:scalarLims} shows the observed and expected $95\%$ CL
model-independent limits on the production cross-section times branching ratio of a narrow-width
scalar resonance, as a function of its mass, in the $WW$ and
$ZZ$ channels combined.  The constraints are compared with the CP-even scalar
singlet model described in Section~\ref{sec:simulation}, for the NDA
and Unsuppressed scenarios. Masses below \limScal~\GeV\ are excluded
for the Unsuppressed scenario.
Figure~\ref{fig:scalarparams} shows the
exclusion contours in the $(c_H/\Lambda,c_3/\Lambda)$ parameter space, derived from
the cross-section limits for three sample masses.
 
\begin{figure}[th!]
  \begin{center}
      \includegraphics[width=0.6\textwidth]{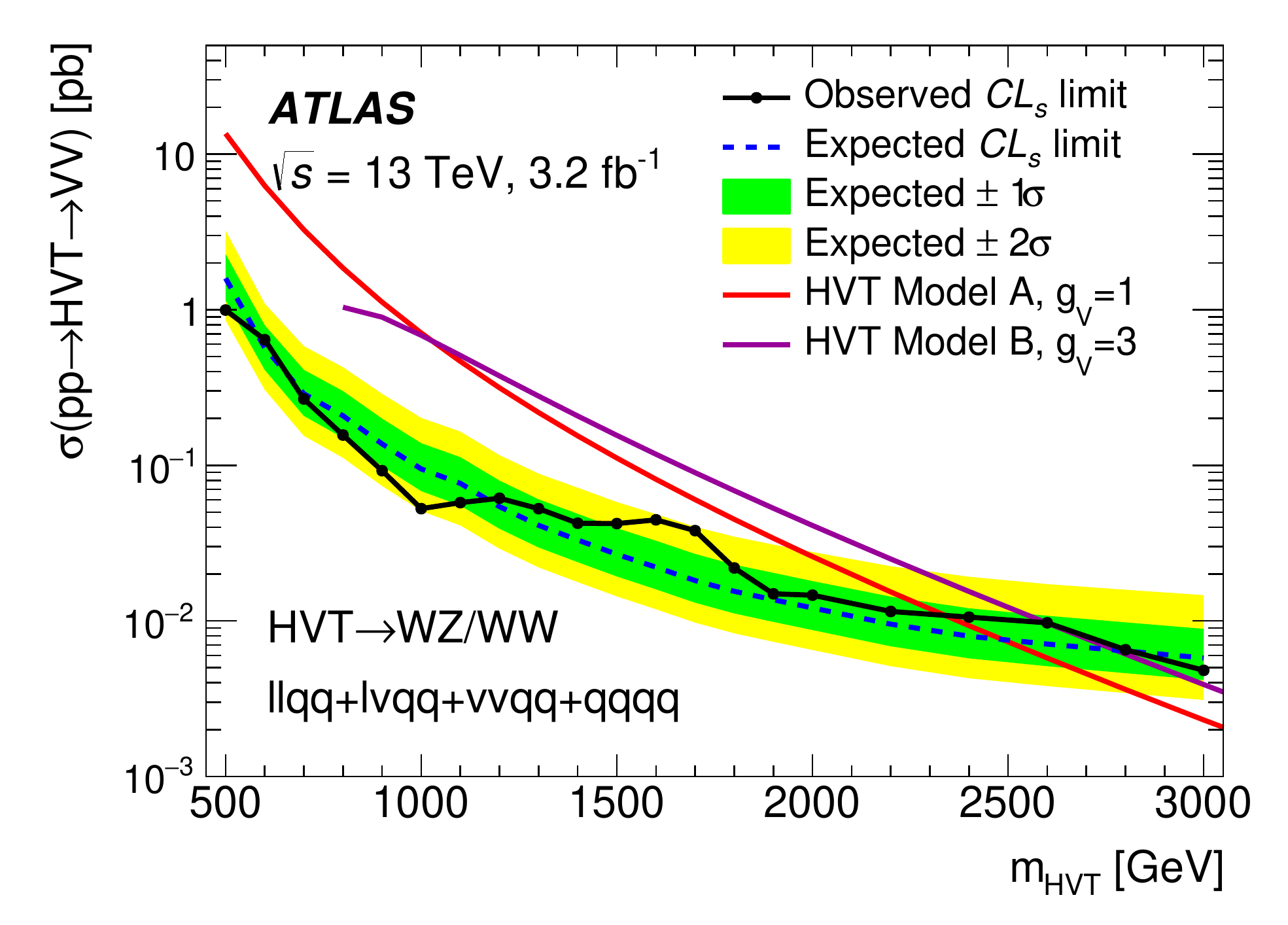}\label{fig:hvt_obs}
  \end{center}
  \vspace{-20 pt}
  \caption{ The observed and expected 95\% CL limits on the cross-section times branching
    ratio to diboson final states for the HVT scenario, compared to the theoretical
    predictions for the HVT model-A with $g_V = 1$ (red line) and
    model-B with $g_V = 3$ (purple line).
  }
  \label{fig:hvtlims}
\end{figure}

\begin{figure}[th!]
  \begin{center}
      \includegraphics[width=0.6\textwidth]{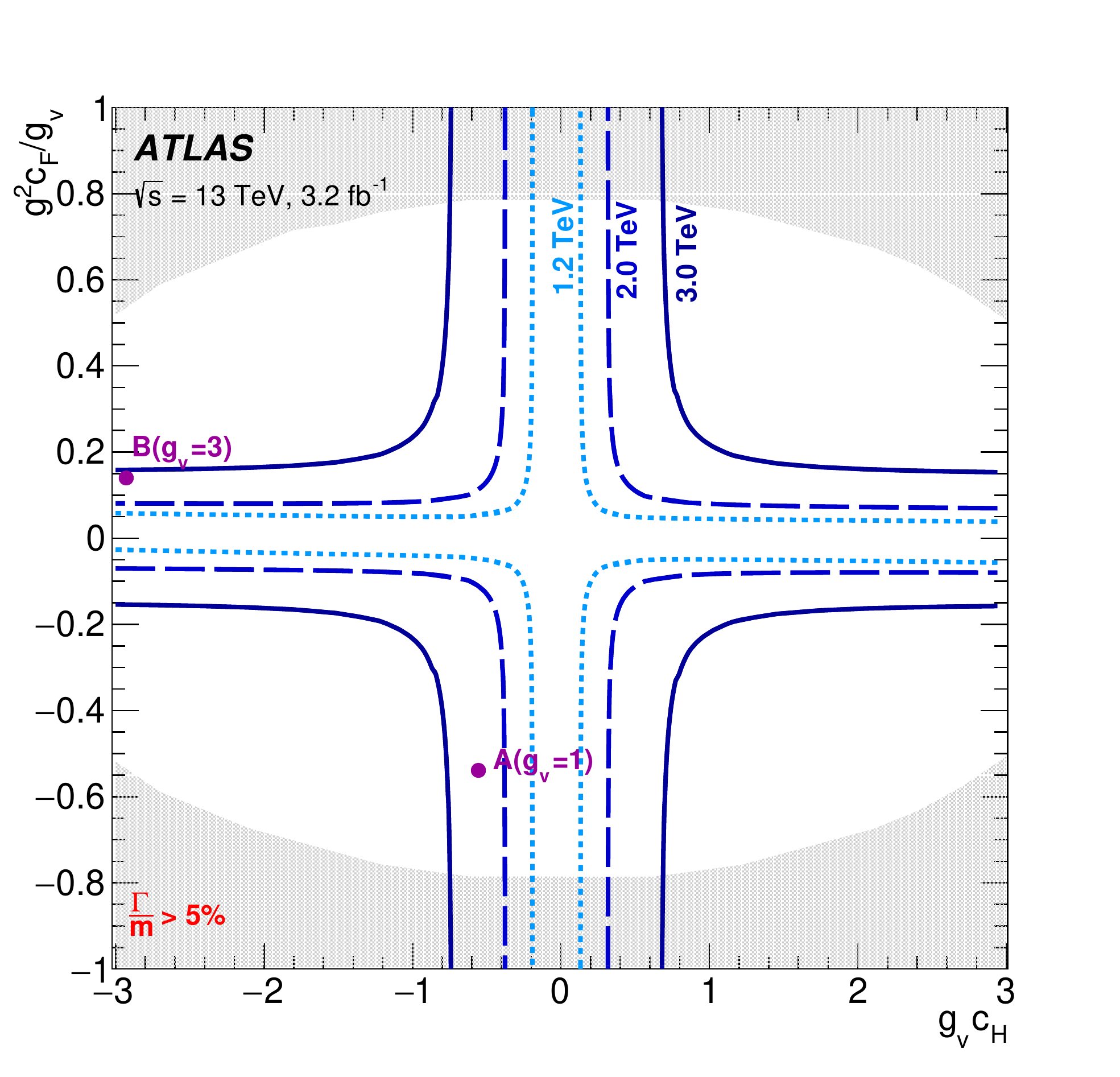}
  \end{center}
  \vspace{-20 pt}
  \caption{ Observed 95\% CL exclusion contours in the HVT parameter
    space $(g^2c_F/g_V, g_Vc_H)$ for resonances of mass $1\tev$,
    $2\tev$ and $3\tev$. Parameters for the benchmark models-A and -B
    are also shown. The grey area indicates the part of the parameter
    space in which the ratio of the resonance's total width $\Gamma$
    to its mass $m$ is higher than $5\%$, which is comparable to the experimental mass resolution.
  }
  \label{fig:hvtparams}
\end{figure}

Figure~\ref{fig:hvtlims} shows the observed and expected
limits obtained in the search for an HVT decaying to $WW$ or $WZ$
states as a function of the mass of the HVT, compared to the
theoretical predictions for the HVT model A assuming $g_V = 1$
and the HVT model-B assuming $g_V = 3$.
For HVT model-B, new gauge bosons with masses below \limHvt~\gev{} are excluded
at the 95\% CL. Results are also shown in Figure ~\ref{fig:hvtparams}
in terms of exclusion contours in the HVT parameter space
$(g^2c_F/g_V, g_Vc_H)$ for different resonance masses~\cite{hvttools}.

\begin{figure}[th!]
  \begin{center}
      \includegraphics[width=0.6\textwidth]{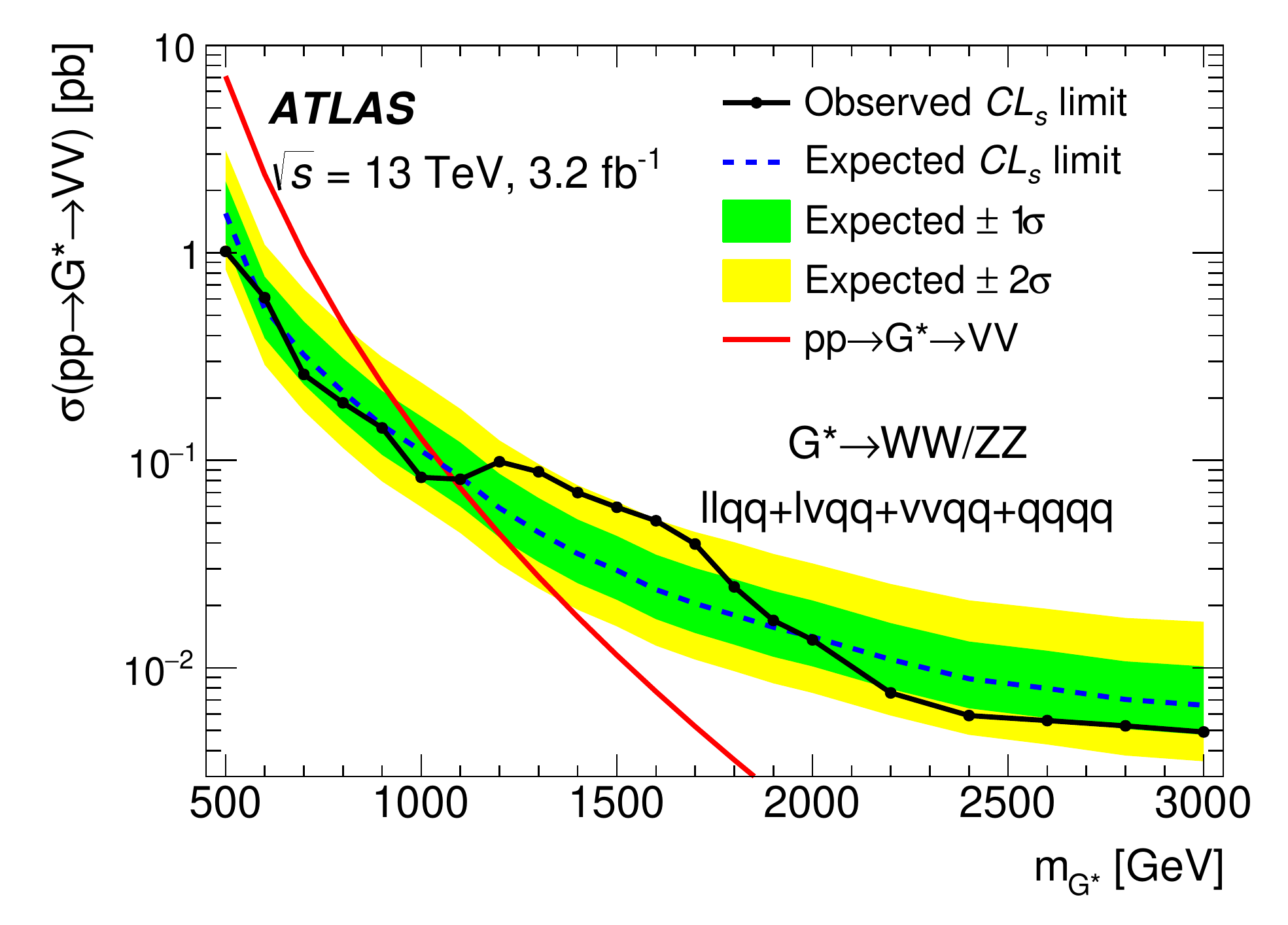}\label{fig:grav_obs}
  \end{center}
  \vspace{-20 pt}
  \caption{ The observed and expected 95\% CL limits on the cross-section times branching
    ratio to diboson final states for bulk RS \grav{}, compared to the
    theoretical prediction for $\kappa/\bar{M}_\text{Pl}=1$ (red line).
  }
  \label{fig:gravlims}
\end{figure}

Similarly, in Figure~\ref{fig:gravlims} the observed and expected
limits obtained in the search for a bulk \grav{} decaying to $WW$ or
$ZZ$ final states are shown and compared to the theoretical prediction
for the bulk RS \grav{} model assuming $\kappa/\bar{M}_\text{Pl} =
1$. Bulk RS \grav{} decaying to $WW$ or $ZZ$ in this model are excluded
if their  mass is below \limGrav~\gev.

No  combination of these results with those at
8~\TeV~\cite{Aad:2015ipg} has been performed. However, the results of the $8\tev$ data analysis in the EGM $W'$ model
have been  used to estimate its sensitivity
to the HVT model~A, including corrections for the  difference in
beam energy and the line-shape of the resonance. This study shows that the current analysis is more sensitive
to the  HVT model~A for triplet masses above 1.6~\TeV, and that the
ratio of the expected cross-section limit to the theoretical
cross-section improves by a factor two for triplet
masses of 2~\TeV.

\FloatBarrier

\section{Conclusion}
\label{sec:conclusion}
A search is performed for resonant $WW$, $WZ$ or $ZZ$
production in final states with at least one hadronically decaying
vector boson, using 3.2~$\ifb$ of proton--proton collisions at
 $\sqrt{s}=13\tev$ recorded in 2015  by the ATLAS detector at the
LHC. No significant excesses are found 
in data compared to the SM predictions. Limits on the production
cross-section times branching ratio into vector-boson pairs are obtained as
a function of the resonance mass for resonances arising from a
model predicting the existence of a new heavy scalar singlet, from a simplified
model predicting a heavy vector-boson triplet, or from a bulk Randall--Sundrum
model with a heavy spin-2 graviton. A scalar resonance with mass below \limScal~\GeV{} predicted by the Unsuppressed model, a heavy vector-boson triplet predicted by model-B with $g_v = 3$ of the HVT parameterisation with mass
below $\limHvt\gev$, and a graviton in the bulk Randall--Sundrum model ($\kappa / \bar{M}_{\mathrm{Pl}} = 1)$
with mass below $\limGrav\gev$
are excluded at the 95\%
confidence level. Limits are also expressed in terms of the parameters
characterising the simplified models considered.

\section*{Acknowledgements}

We thank CERN for the very successful operation of the LHC, as well as the
support staff from our institutions without whom ATLAS could not be
operated efficiently.

We acknowledge the support of ANPCyT, Argentina; YerPhI, Armenia; ARC, Australia; BMWFW and FWF, Austria; ANAS, Azerbaijan; SSTC, Belarus; CNPq and FAPESP, Brazil; NSERC, NRC and CFI, Canada; CERN; CONICYT, Chile; CAS, MOST and NSFC, China; COLCIENCIAS, Colombia; MSMT CR, MPO CR and VSC CR, Czech Republic; DNRF and DNSRC, Denmark; IN2P3-CNRS, CEA-DSM/IRFU, France; GNSF, Georgia; BMBF, HGF, and MPG, Germany; GSRT, Greece; RGC, Hong Kong SAR, China; ISF, I-CORE and Benoziyo Center, Israel; INFN, Italy; MEXT and JSPS, Japan; CNRST, Morocco; FOM and NWO, Netherlands; RCN, Norway; MNiSW and NCN, Poland; FCT, Portugal; MNE/IFA, Romania; MES of Russia and NRC KI, Russian Federation; JINR; MESTD, Serbia; MSSR, Slovakia; ARRS and MIZ\v{S}, Slovenia; DST/NRF, South Africa; MINECO, Spain; SRC and Wallenberg Foundation, Sweden; SERI, SNSF and Cantons of Bern and Geneva, Switzerland; MOST, Taiwan; TAEK, Turkey; STFC, United Kingdom; DOE and NSF, United States of America. In addition, individual groups and members have received support from BCKDF, the Canada Council, CANARIE, CRC, Compute Canada, FQRNT, and the Ontario Innovation Trust, Canada; EPLANET, ERC, FP7, Horizon 2020 and Marie Sk{\l}odowska-Curie Actions, European Union; Investissements d'Avenir Labex and Idex, ANR, R{\'e}gion Auvergne and Fondation Partager le Savoir, France; DFG and AvH Foundation, Germany; Herakleitos, Thales and Aristeia programmes co-financed by EU-ESF and the Greek NSRF; BSF, GIF and Minerva, Israel; BRF, Norway; Generalitat de Catalunya, Generalitat Valenciana, Spain; the Royal Society and Leverhulme Trust, United Kingdom.

The crucial computing support from all WLCG partners is acknowledged gratefully, in particular from CERN, the ATLAS Tier-1 facilities at TRIUMF (Canada), NDGF (Denmark, Norway, Sweden), CC-IN2P3 (France), KIT/GridKA (Germany), INFN-CNAF (Italy), NL-T1 (Netherlands), PIC (Spain), ASGC (Taiwan), RAL (UK) and BNL (USA), the Tier-2 facilities worldwide and large non-WLCG resource providers. Major contributors of computing resources are listed in Ref.~\cite{ATL-GEN-PUB-2016-002}.

\clearpage

\printbibliography

\newpage
\begin{flushleft}
{\Large The ATLAS Collaboration}

\bigskip

M.~Aaboud$^{\rm 135d}$,
G.~Aad$^{\rm 86}$,
B.~Abbott$^{\rm 113}$,
J.~Abdallah$^{\rm 64}$,
O.~Abdinov$^{\rm 12}$,
B.~Abeloos$^{\rm 117}$,
R.~Aben$^{\rm 107}$,
O.S.~AbouZeid$^{\rm 137}$,
N.L.~Abraham$^{\rm 149}$,
H.~Abramowicz$^{\rm 153}$
% ...
\bigskip
\\
% ...
$^{12}$ Institute of Physics, Azerbaijan Academy of Sciences, Baku, Azerbaijan\\
% ...
$^{64}$ University of Iowa, Iowa City IA, United States of America\\
% ...
$^{86}$ CPPM, Aix-Marseille Universit{\'e} and CNRS/IN2P3, Marseille, France\\
% ...
$^{107}$ Nikhef National Institute for Subatomic Physics and University of Amsterdam, Amsterdam, Netherlands\\
% ...
$^{113}$ Homer L. Dodge Department of Physics and Astronomy, University of Oklahoma, Norman OK, United States of America\\
% ...
$^{117}$ LAL, Univ. Paris-Sud, CNRS/IN2P3, Universit{\'e} Paris-Saclay, Orsay, France\\
% ...
$^{135}$ $^{(a)}$ Facult{\'e} des Sciences Ain Chock, R{\'e}seau Universitaire de Physique des Hautes Energies - Universit{\'e} Hassan II, Casablanca; $^{(b)}$ Centre National de l'Energie des Sciences Techniques Nucleaires, Rabat; $^{(c)}$ Facult{\'e} des Sciences Semlalia, Universit{\'e} Cadi Ayyad, LPHEA-Marrakech; $^{(d)}$ Facult{\'e} des Sciences, Universit{\'e} Mohamed Premier and LPTPM, Oujda; $^{(e)}$ Facult{\'e} des sciences, Universit{\'e} Mohammed V, Rabat, Morocco\\
% ...
$^{137}$ Santa Cruz Institute for Particle Physics, University of California Santa Cruz, Santa Cruz CA, United States of America\\
% ...
$^{149}$ Department of Physics and Astronomy, University of Sussex, Brighton, United Kingdom\\
% ...
$^{153}$ Raymond and Beverly Sackler School of Physics and Astronomy, Tel Aviv University, Tel Aviv, Israel\\
% ...
$^{a}$ Also at Department of Physics, King's College London, London, United Kingdom\\
$^{b}$ Also at Institute of Physics, Azerbaijan Academy of Sciences, Baku, Azerbaijan\\
$^{c}$ Also at Novosibirsk State University, Novosibirsk, Russia\\
$^{d}$ Also at TRIUMF, Vancouver BC, Canada\\
$^{e}$ Also at Department of Physics {\&} Astronomy, University of Louisville, Louisville, KY, United States of America\\
% ...
\end{flushleft}

\end{document}